%% file: main.tex
\documentclass[conference]{IEEEtran}

\usepackage{cite}

\ifCLASSINFOpdf
\else
\fi

\ifCLASSOPTIONcompsoc
 \usepackage[caption=false,font=normalsize,labelfont=sf,textfont=sf]{subfig}
\else
 \usepackage[caption=false,font=footnotesize]{subfig}
\fi

\newcommand{\name}{\textsf{{MoPE}}\xspace}
\PassOptionsToPackage{table}{xcolor}
\usepackage{xcolor}                 
\usepackage{colortbl}             
\usepackage{booktabs}
\usepackage{amsmath} 
\usepackage{graphicx}
\usepackage{float}
\usepackage{multirow} 
\usepackage{array}
\usepackage{algorithm}
\usepackage{algorithmic}
\usepackage{xspace}
\usepackage{amssymb} 
\usepackage{threeparttable}
\usepackage{tcolorbox}
\usepackage{enumitem}
\usepackage[hidelinks]{hyperref}

\begin{document}

\title{\name: A Mixture of Password Experts for Improving Password Guessing}


\author{
    \IEEEauthorblockN{
        Mingjian Duan\IEEEauthorrefmark{1},
        Ming Xu\IEEEauthorrefmark{2},
        Shenghao Zhang\IEEEauthorrefmark{1},
        Jiaheng Zhang\IEEEauthorrefmark{2}, and
        Weili Han\IEEEauthorrefmark{1}
    }
    \IEEEauthorblockA{\IEEEauthorrefmark{1}Fudan University, Shanghai, China \\
    Email: \{mjduan24, 22210240071, wlhan\}@fudan.edu.cn}
    \IEEEauthorblockA{\IEEEauthorrefmark{2}National University of Singapore, Singapore \\
    Email: \{mingxu, jhzhang\}@nus.edu.sg}
}

\maketitle

\begin{abstract}
\input{abstract}

\end{abstract}

\input{introduction}

\input{background}
\input{preliminaries}

\input{MoPE}
\input{evaluation}

\bibliographystyle{IEEEtran}
\bibliography{bib}

\clearpage
\appendices
\input{appendix}

\end{document}

%% file: abstract.tex
Textual passwords remain a predominant authentication mechanism in web security.
To evaluate their strength, existing research has proposed several data-driven models across various scenarios. 
However, these models generally treat passwords uniformly, 
neglecting the structural differences among passwords.
This typically results in biased training that favors frequent password structural patterns. 
To mitigate the biased training, we argue that passwords, 
as a type of complex short textual data, 
should be processed in a structure-aware manner by identifying their structural patterns and routing them to specialized models accordingly. 
In this paper, we propose \name, 
a Mixture of Password Experts framework, specifically designed 
to leverage the structural patterns in passwords to improve guessing performance. 
Motivated by the observation that passwords with similar structural patterns (e.g., fixed-length numeric strings) 
tend to cluster in high-density regions within the latent space, our \name introduces:
(1) a novel structure-based method for generating specialized expert models;       
(2) a lightweight gate method to select appropriate expert models to output reliable guesses, 
better aligned with the high computational frequency of password guessing tasks. 
Our evaluation shows that \name significantly outperforms 
existing state-of-the-art baselines in both offline and online guessing scenarios, 
achieving up to 38.80\% and 9.27\% improvement in cracking rate, respectively, 
showcasing that \name can effectively exploit the capabilities of data-driven models for password guessing. 
Additionally, we implement a real-time Password Strength Meter (PSM) based on offline \name, 
assisting users in choosing stronger passwords more precisely with millisecond-level response latency.

%% file: introduction.tex
\section{INTRODUCTION}
Passwords remain an essential component of our digital lives, 
serving as a primary layer of protection for personal data—data that is increasingly vulnerable to compromise. 
For decades, there has been ongoing debate about 
whether we should abandon passwords in favor of more advanced authentication methods, 
such as biometrics or hardware-based solutions~\cite{bonneau:birthday, DBLP:conf/uss/LyastaniSF0B18:password-managers}. 
However, this perspective often overlooks the vulnerabilities inherent in these newer techniques. 
Moreover, relying exclusively on such methods risks creating inequities, 
as they may be more accessible to well-resourced populations 
while excluding those who lack the financial means or institutional support to adopt them. 
Therefore, passwords are not dead—they are still evolving and continue to serve as a reliable backup mechanism for security verification~\cite{yan:password, DBLP:conf/sp/BonneauHOS12, DBLP7:journals/cacm/BonneauHOS15}, 
especially when biometrics or hardware-based authentication methods fail.

Human-created passwords often exhibit predictable patterns due to user habits, making them vulnerable to various attacks. To evaluate password strength, researchers commonly use data-driven models trained on leaked password datasets~\cite{DBLP:conf/ccs/ThomasLZBRIMCEM17:password-map-emails} to uncover latent behaviors in password creation. These models can be effectively deployed as password strength meters (PSMs) in real-world web-based systems. Password strength is typically evaluated within 
offline~\cite{DBLP:conf/sp/BlockiHZ18} 
and online~\cite{DBLP:conf/ccs/WangZWYH16, targeted-guessing-know} guessing scenarios. 
Offline guessing assumes attackers have obtained the hashed password file, 
enabling them to generate a large number of password candidates in descending orders and compare their hashes against the target locally.
The number of offline guesses can be up to $10^{12}$~\cite{DBLP:conf/sp/BlockiHZ18, kelley:guess}, 
depending on computational resources and attack objectives. 
In contrast, online guessing typically assumes a limited number of attempts (e.g., less than 1,000 guesses) due to online rate-limiting mechanisms~\cite{yanq:password, DBLP:conf/ccs/WangZWYH16, DBLP:conf/sp/BonneauHOS12}. 
Given prior knowledge—such as a leaked password from a specific user—attackers may attempt to compromise other accounts owned by the same user by exploiting password reuse behaviors~\cite{han:tdsc/passwordreuse, Das:passwordreuse}. In such cases, models aim to learn password transformation rules (e.g., appending ``!'' to the original password~\cite{Ur:15add!}).

Existing data-driven models mainly include Probabilistic Context-free Grammar (PCFG)~\cite{WeirPCFG}, 
Markov~\cite{Ma:, narayanan:fast, DBLP:journals/tsp/Katz87:backoff, OMEN}, 
FLA~\cite{William:LSTM} for offline guessing,
and TarGuess-II~\cite{DBLP:conf/ccs/WangZWYH16}, Pass2path~\cite{DBLP:conf/sp/PalD0R19:similarity}, 
PassBERT~\cite{xu-real-world-guessing} for online guessing. 
However, all these models treat passwords as a whole during training and inference, 
neglecting nuanced structural differences, 
which leads to biased training toward frequent structural patterns in the datasets.
For example, when the training samples are dominated by a specific structural pattern, e.g., 
the structure of \emph{digit}, 
existing models tend to follow this pattern, assigning higher priority to learn such features.
As a result, they neglect and even fail to process the passwords of other structural patterns such as \emph{letter}, negatively influencing the model's robustness.  
Target passwords often exhibit structural diversity and complexity~\cite{regional-patterns, han:tdsc/passwordreuse, Ur:PGS}, which data-driven guessing frameworks should account for this intrinsic diversity to improve their specificity and robustness.

To overcome this limitation, we argue that passwords should be handled separately rather than collectively modeled. 
Inspired by the MoE architecture~\cite{jacobs1991adaptive, shazeer2017outrageously, eigen2013learning, ma2018modeling}, 
which has the potential to effectively handle 
heterogeneous inputs by dynamically activating specialized experts~\cite{du2022glam, fedus2022switch}, we propose \name, a mixture of password experts. 
\name explicitly partitions the password space based on structural patterns and assigns each subset to a dedicated expert model. The key idea is to leverage multiple specialized models, each focusing on distinct password structures, to better capture the latent variability and enable more fine-grained distribution modeling.    
Several benefits can be introduced in this way: 
(1) it enables specialization, allowing each expert model to focus on a distinct structural pattern, 
such as passwords composed solely of \emph{letter} or \emph{digit}, thereby alleviating model's learning burden on each structural pattern. 
(2) it achieves stronger generalization by effectively capturing diverse and uncommon password structures 
especially when the target passwords differ significantly in structural patterns from the training data.

To build a structure-aware password guessing framework, several challenges should be tackled:
(1)
\textbf{Creating expert models during training}: 
Each expert model should specialize in a distinct, non-overlapping subset of the password space, which requires an effective partitioning of the space into well-defined clusters with significant structural differences. Common approaches~\cite{William:LSTM, yang2024rankguess, pasquini2020improvingpasswordguessingrepresentation} typically use neural networks to extract high-dimensional feature vectors from passwords, which are effective for learning but difficult to delineate clean cluster boundaries, resulting in ambiguous clusters among expert models.  
(2) \textbf{Selecting expert models during inference}:
Based on the input context, \name requires a gate component to
    dynamically invoke the relevant expert models to specifically handle the input case by case. 
    For example, a case with the prefix of ``123456xxx'' should invoke the \emph{digit} related expert models 
    to realize the potential of model effectiveness. However, existing gate methods, 
    usually based on neural networks~\cite{jacobs1991adaptive, shazeer2017outrageously, eigen2013learning, ma2018modeling}, 
    are computationally expensive, complex to train, and slow to execute, 
    especially when dealing with large volumes of passwords in real-world scenarios. 
    Therefore, \name's gate should be both lightweight and accurate, 
    minimizing computational overhead while ensuring proper model selection for optimal performance.
(3) \textbf{Practical deployment}: Integrating multiple expert models and the gate method into real-time applications, 
such as PSMs, requires a sophisticated set to handle the complexity of expert model selection and password predictions. 
To be usable in practice, the system must be specifically optimized to ensure it remains lightweight and suitable for deployment in real-world PSMs.

To tackle these challenges, our \name specifically involves the following components: 
1) to create expert models, we use interpretable structural features like length and character composition, 
applying a password-specific clustering module to group passwords with similar structures. 
To enable specified modeling for a distinct structural pattern, we first pre-train a general-purpose password model, 
and then fine-tune the pre-trained model on each cluster to generate the expert models.  
2) To select only a few highly relevant experts, 
we propose a lightweight center-distance-based gate method as an efficient alternative, assigning weights to expert models based on the distance between the cluster centers and the inputs with a sparse activation scheme.
3) To enable \name's real-time PSM deployment,
we utilize a distilled version model of \name,
ensuring it can handle a password query with millisecond-level response latency without compromising accuracy.

\noindent\textbf{Result.} 
We conduct comprehensive evaluation in offline guessing and online guessing scenarios.
In offline guessing scenario, our \name outperforms state-of-the-art (SOTA) FLA~\cite{William:LSTM} by 0.52\%$\sim$38.80\% (avg. 11.17\%). 
In online guessing scenario, our \name outperforms the SOTA PassBERT~\cite{xu-real-world-guessing} by 5.49\%$\sim$9.27\% (avg. 5.77\%) in 100 and 1,000 guesses. 
Based on our offline \name, 
we further develop a real-time PSM~\cite{DBLP:conf/uss/UrKKLMMPSVBCC12:ur-usenix12-meter} based on distillation techniques, 
achieving 20$\times$ lower query latency than the original offline \name and maintaining accuracy measurement.

We summarize our main contributions as follows:
\begin{itemize}
    \item 
        We propose a novel password modeling framework \name, 
        which includes expert model generation and lightweight gate method 
        to capture and utilize the variations of structural patterns in passwords, 
        thereby enhancing guessing performance.
    \item
        We introduce a distilled version of \name for real-time use as a password strength meter, 
        providing more precise strength ratings to enhance users' risk awareness.
\end{itemize}

\smallskip
\noindent 
We will open-source the necessary code to the community for reproduction.
All releases will be accompanied by strict licensing and ethical-use guidance to ensure responsible use in password security research.

%% file: background.tex
\section{BACKGROUND}
\label{sec:background}

\subsection{Password Guessing Attacks}
\label{subsec:password_guessing_attacks}
To approximately estimate password strength, researchers typically simulate password guessing attacks~\cite{bonneau:guessingyahoo, DBLP16:conf/lisa/FlorencioHO14, yan:password}, which are mainly categorized into offline and online attacks depending on available resources and constraints.

\noindent\textbf{Offline Password Guessing Attacks.}
Early offline password attacks rely on brute-force and rule-based methods, 
which are based on testing all possible combinations of characters 
and possible transformations to dictionary words~\cite{hashcat, John}. 
These approaches always struggle to attack longer or more complex passwords. 
With the increasing availability of leaked password datasets, 
data-driven probabilistic models have become a dominant approach for enhancing password guessing attacks. 
Offline password guessing attacks typically have larger guesses (e.g., $10^{12}$) 
to crack available password hashes leaked from a compromised system. 
Typical offline guessing models include Probabilistic Context-Free Grammars (PCFGs)~\cite{WeirPCFG, modifyPCFG, DBLP:journals/tifs/transPCFG21} 
and Markov-based models~\cite{NarayananMarkov, DBLP:journals/tsp/Katz87:backoff, DBLP:conf/acl-codeswitch/KingBGKWMR14:wordlevel, OMEN}, and neural-network-based models~\cite{William:LSTM, DBLP:conf/acns/HitajGAP19:GAN, yang2024rankguess}. 

\noindent\textbf{Online Password Guessing Attacks.}
Online password guessing typically generates 
password candidates for a target account by leveraging available user-related information as contextual input~\cite{Das:passwordreuse, DBLP:conf/ccs/WangZWYH16}.  
Such information may include previously leaked passwords, personally identifiable information (PII), 
or other attributes related to the user~\cite{DBLP:conf/ccs/WangZWYH16}.  
Different from offline guessing attacks, 
online guessing attacks require interactions with the online system~\cite{florencio:www2007}. 
Consequently, online guessing attacks can be constrained to a limited number of guessing attempts due to rate limiting or account lockout policies.
The earliest forms of online password guessing attacks relied heavily on manual guessing or automated scripts~\cite{morris1979password, klein1990foiling}.
In recent years, researchers have developed various data-driven models to enhance the efficiency of online password guessing, 
such as TarGuess-II~\cite{DBLP:conf/ccs/WangZWYH16}, Pass2path~\cite{DBLP:conf/sp/PalD0R19:similarity} and PassBERT~\cite{xu-real-world-guessing}, which aim to infer another password of a user based on a known one.

\subsection{A Motivating Example~\label{sec:motivating}}
Due to the factors such as cultural differences and personal habits, 
password samples 
exhibit diverse and non-stationary nature~\cite{regional-patterns, Li14, Ur:PGS}. 
Passwords range from highly structured types, such as those consisting entirely of numbers or letters, 
to nearly random types with diverse characters.
This raises the question of whether the processing capabilities of these models are limited when trained on a diverse password dataset, leading to decreased performance~\cite{Ur:PGS}.  

\begin{figure}[h]
    \centering
    \includegraphics[width=0.8\columnwidth]{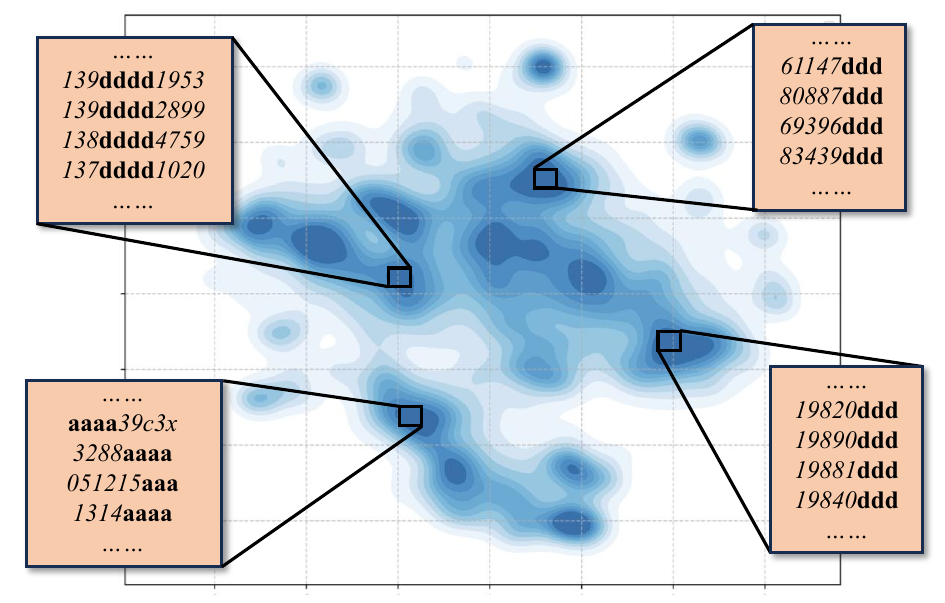}
    \caption{
    The t-SNE visualization of latent space on randomly sampled 10,000 deduplicated passwords. 
    The color intensity indicates the density of password distribution in different regions: \textbf{passwords exhibit cluster features, 
    where each cluster shows their tendencies towards several structures}. 
    Due to the ethical requirement, 
    sampled passwords from dense regions are shown with partial bold masking (lowercase letters $\rightarrow$ \textbf{a}, digits $\rightarrow$ \textbf{d}).
    }
    \label{fig:2}
\end{figure}

Pasquini et al.~\cite{pasquini2020improvingpasswordguessingrepresentation} first suggested that passwords exhibit weak locality, 
referring to the similarities in broad features such as password length, \emph{digit}-to-\emph{character} ratio, 
and the general structure of the password (e.g., presence of \emph{letter}, \emph{digit}, or \emph{special character}). 
Motivated by this, we utilize a pre-trained neural network-based FLA~\cite{William:LSTM} model to process passwords, 
generating 256-dimensional feature vectors, 
and then use a two-dimensional visualization technique 
to show their distribution in the latent space. 
As depicted in Figure~\ref{fig:2}, passwords display obvious cluster structures in the space: 
passwords like ``john1234'' and ``mike5678'' may differ in specific characters, 
but they share common structural traits such as being 8 characters long and having a 4-digit numeric suffix (i.e., weak locality).  
Furthermore, we observe that passwords exhibit \textbf{high-density tendencies} 
within the weak locality.
For example, within the fixed length of 6, 
passwords composed entirely of digits (e.g., ``123456'') 
generally appear more frequently than 
those composed entirely of uppercase letters (e.g., ``ABCDEF''), 
and significantly more than those composed entirely of special characters (e.g., ``!@!@!@'').

Existing studies only touch the surface of structural features, leaving a gap on how to exploit such clusters for improving password guessing. This challenge stems from transferring the structural partitions 
derived from known passwords to make better predictions on a target password set where the partitions are unknown.
We attempt to address this by estimating possible cluster types based on contextual information, 
enabling us to selectively utilize models specialized for those types, 
thereby improving overall model performance.

\subsection{Mixture of Experts}
\label{subsec:moe}
The Mixture of Experts architecture (MoE, for short) is designed 
to handle diverse input patterns by employing a gate network 
that dynamically selects expert models, 
enhancing performance across multiple patterns~\cite{jacobs1991adaptive, shazeer2017outrageously, eigen2013learning, ma2018modeling}.
MoE has gained widespread attention in fields such as natural language processing, 
computer vision and multi-task learning~\cite{riquelme2021scaling, lepikhin2020gshard}, 
due to its effectiveness on processing non-stationary feature distributions. 
By the collaboration of multiple expert models with each focusing on a specific pattern, 
MoE can significantly improve the model performance. 
Due to the different structural clustering of passwords in high dimensional space 
mentioned in Section~\ref{sec:motivating}, 
we infer that password modeling can also benefit from the idea of MoE. 
For instance, one expert might focus on numeric passwords, 
while another handles mixed-character patterns, 
enabling the model to adapt more effectively to diverse password variations.

The core components of MoE include a set of specialized expert models 
and a gate network that assigns weights to each expert based on the input~\cite{jacobs1991adaptive, shazeer2017outrageously, ma2018modeling}. 
During the inference, MoE combines the outputs of the expert models 
based on the weights generated by the gate network. 
Typically, the expert model that specializes in the input 
will gain a relatively higher weight than other expert models, 
thereby achieving specificity. 
Based on the high-density tendencies of passwords, 
we are motivated to propose an architecture that captures these characteristics for improved password guessing. 
Using the current context (e.g., password prefixes or leaked passwords), 
we can select expert models suited to that context to handle relevant patterns, 
adjust their activation levels based on relevance, 
and combine their outputs for the final output.

%% file: preliminaries.tex
\subsection{Password Datasets}
\label{subsec:password_datasets}
In recent years, numerous password leakage incidents have resulted in 
large-scale password datasets becoming publicly accessible~\cite{dell:password, Li14, Das:passwordreuse}. 
These datasets are from various platforms, 
differing in their compositions and distributions. 
Some datasets only contain passwords, 
while others include additional personal information, such as emails or usernames. 
We select suitable publicly available datasets for offline and online guessing experiments.

\noindent\textbf{Offline Guessing Dataset}.
Offline password guessing aims to uncover the hidden distributions
without any additional information. 
Thus, we select publicly available datasets containing only password samples 
as summarized in Table~\ref{tab:offline_password_summary}. 
For English-language password datasets, we use \texttt{RockYou}~\cite{rockyou}, \texttt{Neopets}~\cite{neopets}, 
\texttt{Cit0day}~\cite{cit0day}, 
and \texttt{Rockyou2024}~\cite{rockyou2024}.
Notably, \texttt{Rockyou2024} is the latest dataset, 
which contains mostly rare and uncommon passwords that resemble garbled strings, 
reflecting significant progress of password management tools and user password habits. 
For Chinese-language datasets, 
we use \texttt{CSDN}~\cite{csdn}, 
\texttt{178}~\cite{game178}, 
and \texttt{Taobao}~\cite{taobao}.

\begin{table}[htbp] 
\setlength{\abovecaptionskip}{0pt}
\setlength{\belowcaptionskip}{0pt}
  \caption{Summary of offline password datasets used in this paper.}
  \label{tab:offline_password_summary}
  \centering
  \footnotesize
  \renewcommand\tabcolsep{4.2pt} 
  \begin{tabular}{
    >{\centering\arraybackslash}p{0.5cm}
    >{\centering\arraybackslash}p{2.3cm}
    >{\centering\arraybackslash}p{0.5cm}
    >{\centering\arraybackslash}p{1.2cm}
    >{\centering\arraybackslash}p{2.5cm}
  }
    \toprule
    \textbf{Language} & \textbf{Dataset} & \textbf{Year} & \textbf{Accounts} & \textbf{Service} \\
    \midrule
    \multirow{4}{*}{English}
      & \texttt{Rockyou}        & 2009 & 32,582,532  & Social forum \\
      & \texttt{Neopets}    & 2016 & 67,672,205  & Virtual services \\
      & \texttt{Cit0day}        & 2020 & 86,835,796  & Mixed services \\
      & \texttt{Rockyou2024}    & 2024 & 4,997,274   & Social forum \\
      & *Total (EN)    & —    & 192,087,807 & — \\
    \midrule
    \multirow{4}{*}{Chinese}
      & \texttt{CSDN}         & 2011 & 6,422,884   & Programmer forum \\
      & \texttt{178}          & 2011 & 9,071,979   & Gaming site \\
      & \texttt{Taobao}        & 2016 & 15,073,107  & E-commerce \\
      & *Total (CN)    & —    & 30,567,970  & — \\
    \midrule
    \textbf{Total} & — & — & \textbf{222,655,777} & — \\
    \bottomrule
  \end{tabular}
\end{table}

\noindent\textbf{Online Guessing Dataset}.
Online password guessing aims to identify the transformations between multiple passwords owned by the same user, 
which requires datasets containing user identifiers (e.g., emails).
As shown in Table~\ref{tab:online_password_summary}, 
\texttt{BreachCompilation(4iQ)}~\cite{breachcompilation} is a collection of several well-known websites including Twitter, LinkedIn, Neopets, 
and \texttt{Collection\#1}~\cite{collection1} contains emails with the corresponding passwords leaked in 2019. 
We use emails as the primary key to identify the ownership of passwords, 
locating the leaked passwords belonging to the same user across different platforms. The statistical data in Table~\ref{tab:online_password_summary} shows that each user is associated with an average of more than three leaked passwords across multiple platforms. 
We pair the passwords owned by a user to form source-target password pairs \((p_{\text{src}},p_{\text{tgt}})\). 
The source password represents the password leaked on one platform, 
while the target password corresponds to the password leaked on another platform by the same user. 
Models in the online scenario are expected to generate password candidates based on the leaked password \(p_{\text{src}}\), 
aiming to crack the target password within fewer guesses. 

\begin{table}[htbp]
\setlength{\abovecaptionskip}{0pt}
\setlength{\belowcaptionskip}{0pt} 
  \caption{Summary of online password datasets used in this paper.}
  \label{tab:online_password_summary}
  \centering
  \footnotesize
  \resizebox{\columnwidth}{!}{
  \begin{tabular}{c|c|c}
    \toprule
    & \textbf{\texttt{BreachCompilation (4iQ)}} & \textbf{\texttt{Collection\#1}} \\
    \midrule
    \textbf{Year} & 2017 & 2019 \\
    \textbf{Accounts} & 147,284,401 & 109,191,685 \\
    \textbf{Passwords} & 373,820,141 & 365,336,365 \\
    \midrule
    \multicolumn{1}{c|}{\textbf{Passwords per user}} & & \\
    \quad$\leq$ 10 & 99.3\% & 97.1\% \\
    \quad$>$ 10 & 0.7\% & 2.9\% \\
    \midrule
    \textbf{Password reuse rate} & 4.2\% & 0 \\
    \midrule
    \multicolumn{1}{c|}{\textbf{Edit distance}} & & \\
    \quad$\leq$ 4 & 21.8\% & 28.1\% \\
    \quad$>$ 4 & 78.2\% & 71.9\% \\
    \bottomrule
  \end{tabular}
  }
\end{table}

\noindent\textbf{Data Cleaning and Preprocessing}. 
Following common cleanup processes~\cite{DBLP:conf/sp/PalD0R19:similarity, xu2021chunk, pasquini2021reducing}, 
we filter out invalid passwords. 
Passwords are limited to a maximum of 16 characters~\cite{DBLP:journals/iacr/PasquiniGABC19:representation-learning}, 
ensuring the data quality for our analysis.

\noindent\textbf{Ethical Statement of Datasets}. 
This research aims to enhance password security 
by analyzing publicly accessible datasets.
We strictly follow ethical guidelines, 
focusing on privacy protection and ensuring these datasets are not stored, 
redistributed, or exploited beyond any necessary research purposes.

%% file: MoPE.tex
\begin{figure*}[h]
\setlength{\abovecaptionskip}{0pt}
\setlength{\belowcaptionskip}{0pt}  
    \centering
    \includegraphics[width=2\columnwidth]{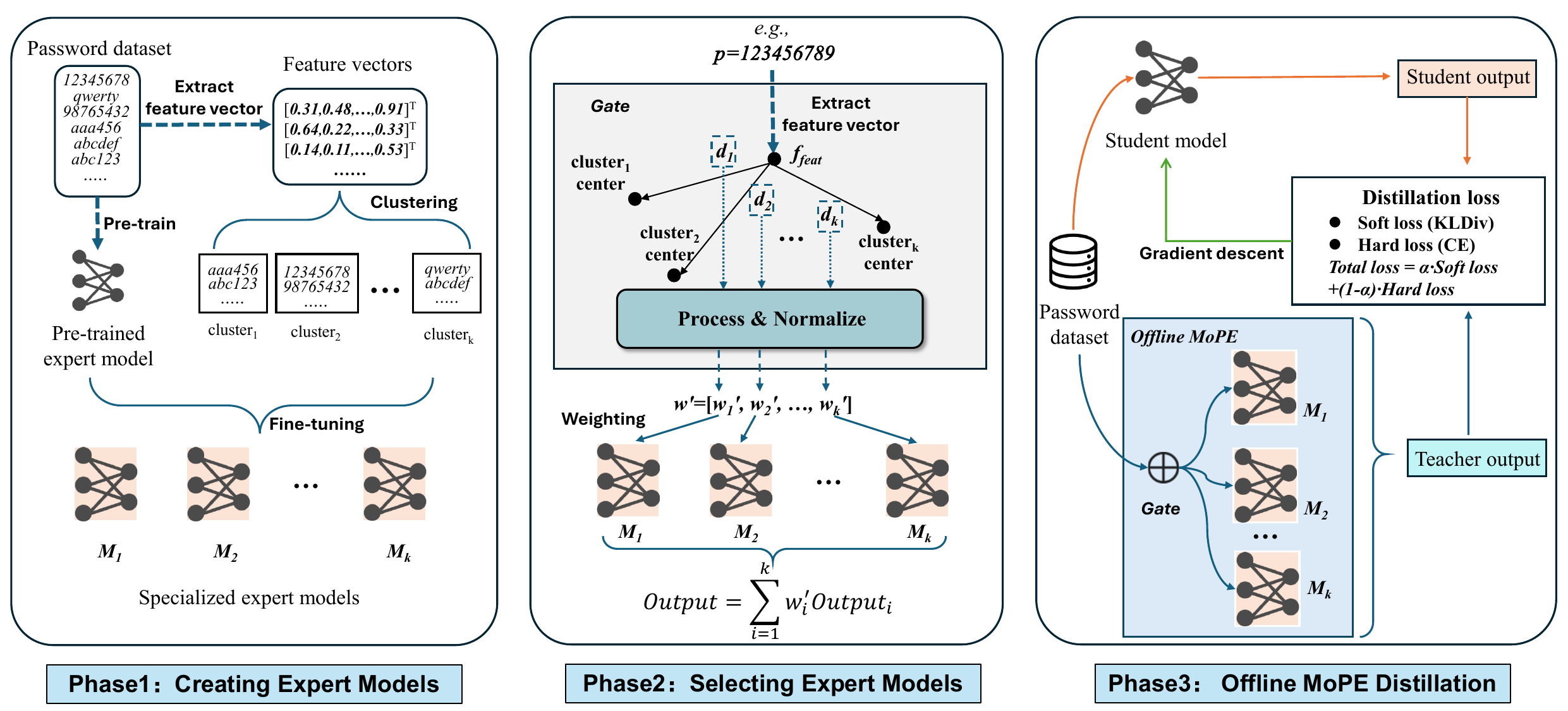}
    \caption{
       Overview of our \name framework for password guessing, which can be tailored into offline and online scenarios. 
    }
    \label{fig:5}
\end{figure*}

\section{\name FRAMEWORK}
\label{sec:framework}

In this section, we detail our \name framework (shown in Figure~\ref{fig:5}), 
which consists of expert models' creating, selecting and distilling.
First, to create expert models, \name performs password clustering to identify the underlying structural patterns of passwords,
where each cluster is used to fine-tune an expert model. 
Then, \name applies a lightweight gate method with the password or prefix as input, 
enabling \name to select the optimal expert models for inference. 
Finally, to enable \name to be practically and efficiently applied, we use a distillation method to compress offline \name's knowledge into a single expert model.

\subsection{Expert Model Creation}~\label{subsec:clustering}

\subsubsection{Password Clustering} 
Creating expert models requires identifying relevant password clusters. 
Existing neural network-based methods such as FLA~\cite{William:LSTM} can produce password clustering, 
relying on high-dimensional feature representations. 
However, the high-dimensional vectors are inherently unsuitable for clustering due to issues like sparse distribution and the curse of dimensionality, 
suffering from limited interpretability and high computational costs, 
also hindering a clear understanding of the underlying structural patterns in passwords. 
To enable each expert model to learn a specific structural pattern, 
we categorize passwords into high-quality clusters based on their structural characteristics.
We do not consider semantic-based features, because passwords are typically short in length thus generally lack clear semantic structures.
The structure-aware pattern plays an important role in password composition, 
confirmed by a series of works~\cite{WeirPCFG, NarayananMarkov, mayer2017second, komanduri2011passwords}, 
suited for password clustering associated with expert models. 
The cluster quality directly influences the expert model's performance, given that each cluster corresponds to an expert model. 
We quantify the clustering quality in Section~\ref{sec:additional-evaluation}.

Specifically, we design a feature extraction function that maps 
each password \( p \) to a multi-dimensional feature vector \( \mathbf{f}_{feat}(p) = [f_1, f_2, \dots, f_m] \), 
where \( m \) denotes the number of features and $f_i$ is the $i$-th dimensional feature of $p$ for \(i = 1,2,...,m\). We manually select eight key features~\cite{WeirPCFG, NarayananMarkov, mayer2017second, komanduri2011passwords} shown in Table~\ref{tab:password_features}, 
which effectively distinguish password structural patterns
as validated in Section~\ref{sec:evaluation}.
For a given set of passwords \( \mathcal{P} \), 
the feature extraction function \( \mathbf{f}_{feat}: \mathcal{P} \rightarrow \mathbb{R}^m \) is defined 
using these features, thus the feature vector is 
\( \mathbf{f}_{feat}(p) = [l, r_{\mathbb{D}}, r_{\mathbb{L}}, r_{\mathbb{U}}, r_{\mathbb{S}}, s, d_{\text{max}}, a_{\text{max}}] \). 
For example, for the password ``Abc123'', the corresponding
\( \mathbf{f}_{feat}(\textnormal{“Abc123"}) = [6, 0.5, 0.333, 0.167, 0, 1, 3, 3] \).  
We standardize the feature vectors to eliminate bias. 
For each password \( p_i \) (where \( i = 1, 2, \dots, n \)), 
the feature vector \( \mathbf{f}_{feat}(p_i) \in \mathbb{R}^m \). 
Thus, the feature vectors of all \( n \) passwords 
form an \( n \times m \) feature matrix \( \mathbf{F} \):

\begin{table*}[htbp]
\setlength{\abovecaptionskip}{0pt}
\setlength{\belowcaptionskip}{0pt}  
\centering
\caption{Definitions of password structural features used for clustering in \name}
\renewcommand{\arraystretch}{1.2}
\renewcommand\tabcolsep{20pt}
\begin{threeparttable}
\begin{tabular}{ccc}
\toprule
\textbf{Feature} & \textbf{Definition} & \textbf{Example (e.g., \texttt{"abc123"})} \\
\midrule
Length ($l$) & $l = |P|$ & $l = 6$ \\
Digit Ratio ($r_{\mathbb{D}}$) & $r_{\mathbb{D}} = \frac{n_{\mathbb{D}}}{l}$ & $r_{\mathbb{D}} = \frac{3}{6} = 0.5$ \\
Lowercase Ratio ($r_{\mathbb{L}}$) & $r_{\mathbb{L}} = \frac{n_{\mathbb{L}}}{l}$ & $r_{\mathbb{L}} = \frac{3}{6} = 0.5$ \\
Uppercase Ratio ($r_{\mathbb{U}}$)& $r_{\mathbb{U}} = \frac{n_{\mathbb{U}}}{l}$ & $r_{\mathbb{U}} = \frac{1}{6} \approx 0.167$ (for \texttt{"Abc123"}) \\
Special Character Ratio ($r_{\mathbb{S}}$) & $r_{\mathbb{S}} = \frac{n_{\mathbb{S}}}{l}$ & $r_{\mathbb{S}} = \frac{1}{6} \approx 0.167$ (for \texttt{"abc\#12"}) \\
Switch Count ($s$) & $s = \text{number of type transitions in } P$ & $s = 1$ (\emph{letter} $\rightarrow$ \emph{digit}) \\
Max Consecutive Digits ($d_{\max}$) & $d_{\max} = \max_i \{ \text{len of \emph{digit} run}_i \}$ & $d_{\max} = 3$ (for \texttt{"ab12cd345"}) \\
Max Consecutive Letters ($a_{\max}$) & $a_{\max} = \max_i \{ \text{len of \emph{alpha} run}_i \}$ & $a_{\max} = 2$ (for \texttt{"ab12cd"}) \\
\bottomrule
\end{tabular}
\begin{tablenotes}[flushleft]
   \item[$\bullet$] $P$: password string; $l$: length of $P$; $\mathbb{D}, \mathbb{L}, \mathbb{U}, \mathbb{S}$: sets of \emph{digit}, \emph{lowercase}, \emph{uppercase}, and \emph{special character};
   \item[$\bullet$] $n_{\mathbb{D}}, n_{\mathbb{L}}, n_{\mathbb{U}}, n_{\mathbb{S}}$: number of digits, lowercase, uppercase, and special characters; $s$: type-switch count;
    \item[$\bullet$] $d_{\max}, a_{\max}$: maximum consecutive \emph{digit}, \emph{letter} lengths.
\end{tablenotes}
\end{threeparttable}
\label{tab:password_features}
\end{table*}

\begin{equation}\nonumber
\footnotesize
\mathbf{F} = 
\begin{bmatrix}
\mathbf{f}_{feat}(p_1) \\
\mathbf{f}_{feat}(p_2) \\
\vdots \\
\mathbf{f}_{feat}(p_n)
\end{bmatrix}
=
\begin{bmatrix}
f_{11} & f_{12} & \cdots & f_{1m} \\
f_{21} & f_{22} & \cdots & f_{2m} \\
\vdots & \vdots & \ddots & \vdots \\
f_{n1} & f_{n2} & \cdots & f_{nm}
\end{bmatrix}\text{.}
\end{equation}

\noindent For each feature \( j = 1, 2, \dots, m \), 
we compute the mean \( \bar{x}_j \) and standard deviation \( \sigma_j \) across \( n \) passwords. 
Each feature \( f_{ij} \) is then standardized as:

\begin{equation}\nonumber
    f'_{ij} = \frac{f_{ij} - \bar{x}_j}{\sigma_j}, \quad i = 1, 2, \dots, n, \quad j = 1, 2, \dots, m,
\end{equation}
yielding the standardized matrix \( \mathbf{F}' \) with elements \( f'_{ij} \). 
The \( i \)-th row of \( \mathbf{F}' \) is the standardized feature vector \( \mathbf{f}_{feat}'(p_i) \).

\noindent\textbf{Clustering}. 
We apply the K-Means algorithm to the standardized feature vectors, 
as shown in Algorithm~\ref{alg:kmeans}, to generate cluster labels for each password, thereby grouping passwords with the same label into clusters \( \mathcal{C}_i \) (\( i = 1, 2, \dots, k \)) as shown in Figure~\ref{fig:5}.

\begin{algorithm}
\footnotesize
    \caption{Clustering Algorithm}
    \label{alg:kmeans}
    \begin{algorithmic}[1]
    \STATE \textbf{Input:} Standardized feature matrix \(\mathbf{F}'\), number of clusters \(k\)
    \STATE \textbf{Output:} Cluster labels for each sample
    \STATE Randomly select \(k\) samples from \(\mathbf{F}'\) as cluster centers \(\mathbf{\mu}_1, \mathbf{\mu}_2, \dots, \mathbf{\mu}_k\)
    \WHILE{cluster centers have not converged}
    \FOR{each sample \(\mathbf{f}_{feat}'(p_i)\) in \(\mathbf{F}'\)}
    \STATE Compute distances \(d_{ij} = \|\mathbf{f}_{feat}'(p_i) - \mathbf{\mu}_j\|\) for \(j = 1, 2, \dots, k\)
    \STATE Assign \(\mathbf{f}_{feat}'(p_i)\) to cluster \(C_j\) where \(j = \arg\min_j d_{ij}\)
    \ENDFOR
    \FOR{each cluster \(C_j\), \(j = 1, 2, \dots, k\)}
    \STATE \(\mathbf{\mu}_j = \frac{1}{|C_j|} \sum_{\mathbf{f}_{feat}'(p_i) \in C_j} \mathbf{f}_{feat}'(p_i)\)
    \ENDFOR
    \ENDWHILE
    \STATE \textbf{Return} cluster labels for each sample
    \end{algorithmic}
\end{algorithm}

To determine the appropriate number of clusters,
we use the silhouette score~\cite{Rousseeuw1987}, which quantifies cluster cohesion and separation and provides a reliable basis for assessing clustering quality,
as detailed in Appendix~\ref{subsec:sil_define}. 
To balance clustering quality and cluster size, we select the smallest \(k\)  for which the silhouette score first exceeds a given threshold (e.g., 0.5 or 0.7~\cite{wikipedia:silhouette}), ensuring high-quality clusters with sufficient samples.
Formally, we define the optimal \( k^* \) as:
\begin{equation}\nonumber
    k^* = \min \{ k \in K \mid S(k) > \tau \}\text{.}
\end{equation}
Here, \( K = \{2, 3, \dots, k_{\text{max}}\} \) 
represents the range of possible cluster numbers, \(S(k)\) denotes the silhouette score of \(k\) clusters, 
and \( \tau\) is a manually set threshold.

\subsubsection{Expert Model Fine-tuning}
\label{subsec:expert_model_gen}
With the clustered passwords, 
\name employs a two-phase process to produce specialized expert models, 
preserving general knowledge while enabling structure-specific specialization.

\noindent\textbf{Pre-training}.
We initialize a base model \(\mathcal{M}\) with parameters \(\theta\) as the expert model, 
pre-training it on the entire password dataset \(\mathcal{P}\). 
The pre-training objective is defined as minimizing the task-specific loss function \(\mathcal{L}_{\text{pre}}\):

\begin{equation}\nonumber
\theta^* = \arg\min_{\theta} \mathcal{L}_{\text{pre}}(\theta; \mathcal{P}),
\end{equation}
where \(\mathcal{L}_{\text{pre}}\) is customized to the guessing task. 
For instance, in our sequence generation tasks~\cite{William:LSTM}, 
the loss is the negative log-likelihood over prefix-target pairs. 

\noindent\textbf{Fine-tuning}.
Subsequently, we fine-tune the pre-trained model \(\mathcal{M}\) on each cluster \(\mathcal{C}_i\) (\(i = 1, 2, \dots, k\)) 
to generate specialized expert models \(\mathcal{M}_i\) with parameters \(\theta_i\). 
The fine-tuning objective mirrors the pre-training loss but is applied to cluster-specific data: 
\begin{equation}\nonumber
\theta_i^* = \arg\min_{\theta_i} \mathcal{L}_{\text{fine}}(\theta_i; \mathcal{C}_i), \quad \forall i = 1, 2, \dots, k,
\end{equation} 
where \(\mathcal{L}_{\text{fine}}\) adopts the same functional form as \(\mathcal{L}_{\text{pre}}\). 
After fine-tuning, we obtain a set of expert models, each corresponding to one of the clusters. 
For example, if \(k = 50\) clusters are identified, we obtain 50 expert models, 
each optimized for its cluster's structural pattern. 
To mitigate overfitting to limited cluster data while preserving general password knowledge, 
fine-tuning is restricted to a few epochs.

\subsection{Expert Model Selection}
\label{subsec:lightweight_gate}
Given a set of expert models, \name is required to select appropriate expert models in different contexts using a gate method~\cite{shazeer2017outrageously}.
Existing frameworks typically use neural network-based methods 
to determine the suitable expert models,
which are resource-intensive and time-consuming.
To address this, we propose a center-distance-based gate method as our efficient alternative in \name.

Our intuition is that each cluster captures a distinct structural pattern, allowing the cluster center 
\(\mu_j\) of \(C_j\) to serve as its representative feature vector. By measuring the distance between an input's feature vector and 
\(\mu_j\), we assess its structural similarity to the cluster. 
For example, the feature vector of the prefix ``1234'' closely aligns with the center vectors of clusters 
containing data like ``123456'' or ``123456abc'', 
allowing us to quickly identify the expert models based on the feature vector of the prefix and each cluster.  
We then assign weight to each expert model based on sparse activation criteria (i.e., retaining only weights above a threshold).
Suppose the clustering analysis produces \( k \) clusters 
with centers \( \mathbf{\mu}_1, \mathbf{\mu}_2, \dots, \mathbf{\mu}_k \in \mathbb{R}^m \), 
where \( m \) denotes the dimensionality of the feature vectors. 
For an input prefix \( x \), we extract its feature vector \( \mathbf{f}(x) \in \mathbb{R}^m \) 
and standardize it to obtain \( \mathbf{f}'(x) \). 
The Euclidean distance between \( \mathbf{f}'(x) \) and each cluster center \( \mathbf{\mu}_j \) is then computed as:

\begin{equation}\nonumber
d_j = \| \mathbf{f}'(x) - \mathbf{\mu}_j \|_2, \quad j = 1, 2, \dots, k. 
\end{equation}

\noindent To emphasize the relative closeness of the prefix to different clusters, 
we use an exponential decay function to obtain the weights of the expert models, defined as: 

\begin{equation}\nonumber
w_j = \frac{p_j}{\sum_{i=1}^k p_i}, \quad j = 1, 2, \dots, k,  
\end{equation}

\noindent where: 

\begin{equation}\nonumber
p_j = e^{-d_j}, \quad j = 1, 2, \dots, k. 
\end{equation}

\noindent Here \( w_j \) represents the activation weight for the \( j \)-th expert model. 
A larger \( w_j \) indicates higher similarity between the input prefix \( x \) and the \( j \)-th cluster, implying that the corresponding expert model is more appropriate for processing this password.
To further enhance the speed of inference without sacrificing accuracy, 
we deactivate the expert models with weights below a specified threshold. 
Given an input prefix \( x \), the center-distance-based gate method generates a weight vector
\begin{equation}\nonumber
    \mathbf{w} = [w_1, w_2, \ldots, w_k] \in \mathbb{R}^k.
\end{equation}
Weights below a threshold \( \frac{1}{k \cdot \beta} \) are zeroed out, where $\beta$ is a tunable parameter controlling the sparsity level. 
The activated weights are defined as:
\begin{equation}\nonumber
w_j' =
\begin{cases}
w_j, &w_j \ge \frac{1}{k \cdot \beta} \\
0, & \text{otherwise}
\end{cases}
, \quad j = 1, 2, \dots, k. 
\end{equation}
The activated weights are normalized over the selected indices:
\begin{equation}\nonumber
w_j' \leftarrow \frac{w_j'}{\sum_{i=1}^{k} w_i'}, \quad j = 1, 2, \dots, k. 
\end{equation}

\noindent We use \(g\) to denote the center-distance-based gate method, which maps an input prefix \(x\) to a weight vector \(\mathbf{w}' \in \mathbb{R}^k\), 
defined as:
\begin{equation}\nonumber
g(x)= \mathbf{w}'=[w_1', w_2', \ldots, w_k'] \in \mathbb{R}^k. 
\end{equation}

\noindent Passwords can be regarded as special prefixes and thus be processed using \(g\).

\subsection{Offline and Online Guessing Adaptation}
The \name framework is inherently adaptable to diverse password guessing scenarios by enabling expert model creation and selection based on scenario-specific objective functions, with offline and online attacks serving as two representative cases.


\subsubsection{Offline Guessing}
Formally, the goal in the offline scenario is to model a probability distribution \( P(y) \)~\cite{WeirPCFG, NarayananMarkov, William:LSTM} 
over the space of passwords \( y \in \mathcal{Y} \), 
such that high-probability candidates can be efficiently enumerated and tested against leaked password hashes. 
A \textbf{sequence generation} approach~\cite{William:LSTM} is employed to model the password distribution in \name. 
Specifically, we decompose the generation of a password \( y = (c_1, c_2, \dots, c_L) \) into a character-level autoregressive process:

\begin{equation}\nonumber
P(y) = \prod_{j=1}^{L} P(c_j \mid c_1, c_2, \dots, c_{j-1}),
\end{equation}
where each \( c_j \in \Sigma \), \(\Sigma\) is the predefined character set.

The core of sequence generation is predicting the probability distribution of the next character 
based on the current prefix. 
Specifically, we define two special tokens, 
\textsf{PASSWORD\_START} and \textsf{PASSWORD\_END}, 
to control the generation of passwords. 
The model outputs a probability distribution of characters based on the current prefix, with the beginning generation containing only \textsf{PASSWORD\_START}. 
We sample characters and append them to the prefix, 
gradually generating a string until \textsf{PASSWORD\_END} is sampled or the maximum length is reached.

\noindent\textbf{Offline \name Design}.  
To tailor \name into offline password guessing, 
we extract all prefix-target pairs \((x_j, y_j)\) from the sequence \(y\), 
where the prefix \(x_j = (c_0, c_1, \dots, c_{j-1})\) is used to predict the next character \(y_j = c_j\), 
following the training of sequence generation model~\cite{William:LSTM}. 
This transforms each password into a series of training samples 
\(
\{(x_1, y_1),\ (x_2, y_2),\ \dots,\ (x_{L+1}, y_{L+1})\}.
\)
Let \(\mathcal{D}_{off} = \{(x^{(i)}, y^{(i)})\}_{i=1}^{N}\) 
denote the entire set of such training pairs derived from the offline dataset. 
Given a prefix \(x\), the model learns a conditional distribution 
\(P_\theta(c \mid x)\) over the next character \(c \in \Sigma\):  

\begin{equation}\nonumber
\begin{aligned}
\theta^* = \arg\min_{\theta} \; \mathcal{L}_{\text{pre}}(\theta;\mathcal{D}_{off}) 
            = -\sum_{i=1}^{N} \log P_\theta(y^{(i)} \mid x^{(i)}),
\end{aligned}
\end{equation}
where \(\theta\) denotes the trainable parameters of the model. 
This objective encourages the model to assign higher probability 
to the ground-truth next character given its prefix.
We therefore obtain the expert models \(\mathcal{M}_i\) (\(i = 1, 2, \dots, k\)) for the offline scenario. 
Given the current prefix \( x \), 
we obtain the weights \( w_j' \) for each expert model \(\mathcal{M}_j\) by calculating \(g(x)\). 
Let \( P_j(c \mid x) \) denote the probability distribution over the next token \( c \in \Sigma\) given prefix \( x \), 
predicted by the expert model \(\mathcal{M}_j\).  
The final output distribution of offline \name is a weighted sum of the activated expert models' distributions:

\begin{equation}\nonumber
    P_{\text{\name}}(c \mid x) = \sum_{1 \leq j \leq k} w_j' \cdot P_j(c \mid x). 
\end{equation}

\noindent The resulting distribution \( P_{\text{\name}}(c \mid x) \) can be used to sample the next token given the prefix \(x\), 
which we can apply to generate passwords as depicted in Algorithm~\ref{alg:mope_offline}. 

\begin{algorithm}
\footnotesize
    \caption{Offline \name Password Generation}
    \label{alg:mope_offline}
    \begin{algorithmic}[1]
    \STATE \textbf{Input:} expert models \(\{\mathcal{M}_j\}_{j=1}^k\) in \name, $l_{max}$, $l_{min}$, threshold $\tau$
    \STATE \textbf{Output:} Generated password set $\mathcal{P}_{gen}$
    \STATE $\mathcal{P}_{gen} \leftarrow \emptyset$, queue $\mathcal{Q} \leftarrow \{(\texttt{\textless PASSWORD\_START\textgreater}, 1.0)\}$
    \WHILE{$\mathcal{Q} \neq \emptyset$}
        \STATE Pop $(x, prob)$ from $\mathcal{Q}$
        \IF{$|x|\geq l_{max}$ or $x$ ends with \texttt{\textless PASSWORD\_END\textgreater}}
            \IF{$|x|\geq l_{min}$}
                \STATE $\mathcal{P}_{gen} \leftarrow \mathcal{P}_{gen} \cup \{x\}$
            \ENDIF
            \STATE \textbf{continue}
        \ENDIF
        \STATE Calculate $P_{\text{\name}}(c \mid x)$
        \FOR{$c$ where $prob \cdot P_{\text{\name}}(c \mid x) \geq \tau$}
            \STATE $\mathcal{Q} \leftarrow \mathcal{Q} \cup \{(x+c, prob \cdot P_{\text{\name}}(c \mid x))\}$
        \ENDFOR
    \ENDWHILE
    \STATE Sort $\mathcal{P}_{gen}$ by \(prob\) (descending)
    \STATE \textbf{return} $\mathcal{P}_{gen}$
    \end{algorithmic}
\end{algorithm}

\subsubsection{Online Guessing} 
The goal is to model a conditional distribution \(P(p \mid C)\), where \(p\) is the unknown target password and \(C\) represents contextual information associated with the user.
Here, we define the online guessing scenario ~\cite{DBLP:conf/ccs/WangZWYH16, DBLP:conf/sp/PalD0R19:similarity}, 
where the context \(C\) is a leaked password \(p'\) from the same user on another platform: \(P(p \mid C)\) becomes \(P(p \mid p')\). 
This setting reflects a common real-world threat,
where users reuse passwords by applying simple modifications to existing ones, which attackers can exploit across services~\cite{Das:passwordreuse, DBLP:conf/sp/PalD0R19:similarity, DBLP:conf/ccs/WangZWYH16}.

To this end, we aim to learn the transformation paths between passwords, 
which can be applied to the known \( p' \) to yield the target passwords of the same user on another platform. 
The \textbf{edit operation}~\cite{DBLP:conf/sp/PalD0R19:similarity, wang2023pass2edit} is typically employed to construct the transformation paths: 
the transformation between passwords can be represented 
by a sequence of edit operations \(e = (e_1, e_2, \dots, e_n)\).   
Sequential action modeling reflects password reuse behaviors, 
and is widely adopted in online guessing models~\cite{DBLP:conf/sp/PalD0R19:similarity, wang2023pass2edit, xu-real-world-guessing}. 
The order of operations reflects the incremental nature of how users modify their passwords step-by-step. 
With the sequence of edit operations, we can use an edit function \(T(\cdot)\) 
to transform \(p'\) into a password candidate \(\hat{p}\): 
\begin{equation}\nonumber
    \hat{p} = T(p', e),
\end{equation}
where each edit operation \( e_j = (op, pos) \) 
denotes an edit operation \(op\) at the position \(pos\) of the password string. Formally, given a password pair \((p', p)\) from the same user, 
the conditional probability of generating the target password from a leaked password is modeled as:

\begin{equation}\nonumber
P(p \mid p') = P(e \mid p') = \prod_{j=1}^{n} P(e_j \mid p', e_1, \ldots, e_{j-1}).
\end{equation}
We follow the operations of \texttt{del}, \texttt{ins}, \texttt{rep} and \texttt{end}. 
Assuming the maximum password length is \(\ell\), the operations are defined as follows: 
\texttt{del} deletes the character at position \(pos\); 
\texttt{ins} inserts a character \(c\) at position \(pos\); 
\texttt{rep} replaces the character at \(pos\) with a new character \(c\); 
\texttt{end} terminates the editing process, indicating that \(T(p', e)\) is the final output \(\hat{p}\).

\noindent\textbf{Online \name Design}. 
The expert models of online \name model the conditional transformation 
from \( p_{\text{src}} \) to \( p_{\text{tgt}} \) as a 
sequence of atomic edit operations \( e = (e_1, e_2, \dots, e_n) \), 
drawn from a predefined operation space:
\begin{equation}\nonumber
\mathcal{O} = \{\texttt{ins}(c, pos),\ \texttt{del}(pos),\ \texttt{rep}(c, pos),\ \texttt{end} \mid c \in \Sigma\},
\end{equation}
where the transformation paths progressively modify the source password. We construct a supervised training dataset 
\(\mathcal{D}_{\text{on}} = \{(p_{\text{src}}, p_{\text{tgt}})\}\), 
applying an edit distance threshold to ensure learnable transformations.
For each password pair, we compute a minimal transformation path 
\( e = (e_1, \ldots, e_n) \in \mathcal{O}^* \)  as the ground truth. 
The model learns to generate the edit operation sequence conditioned on the source password:
\begin{equation}\nonumber
P(e \mid p_{\text{src}}) = \prod_{j=1}^{n} P(e_j \mid p_{\text{src}}, e_{<j}), 
\end{equation}
where the objective is to maximize the log-likelihood of the correct edit sequence over the training set. Specifically, 
let each training example be denoted as \(d=(p_{\text{src}}, p_{\text{tgt}})\), 
we optimize the model parameters \(\theta\) below:

\begin{equation}\nonumber
\footnotesize
    \begin{aligned}
    \theta^* &= \arg\min_{\theta} \; \mathcal{L}_{\text{pre}}(\theta; \mathcal{D}_{\text{on}}) \\
    &= \arg\min_{\theta} \left( - \frac{1}{|\mathcal{D}_{\text{on}}|} \sum_{d \in \mathcal{D}_{\text{on}}} \log P_\theta(e \mid p_{\text{src}}) \right) \\
    &= \arg\min_{\theta} \left( - \frac{1}{|\mathcal{D}_{\text{on}}|} \sum_{d \in \mathcal{D}_{\text{on}}} \sum_{j=1}^{n} \log P_\theta(e_j \mid p_{\text{src}}, e_1, \dots, e_{j-1}) \right). 
\end{aligned}
\end{equation}

\noindent Following the expert model creation, we obtain the expert models \(\mathcal{M}_i\) (\(i = 1, 2, \dots, k\)) for the online scenario.  

Given the current password \( p' \), 
we can obtain the weights \( w_j' \) to calculate the likelihood of each edit operation for \( p' \) based on each expert model's output.  
Denote that \(P_j(e \mid p')\) is the probability distribution predicted by the \( j \)-th expert model
for the password candidate \(\hat{p}\) transformed from \( p' \) using edit operations \(e\),  
thus the final probability assigned to each candidate \(\hat{p}\) under online \name is computed as:
\begin{equation}\nonumber
\begin{aligned}
P_{\text{\name}}(\hat{p}\mid p')=P_{\text{\name}}(e \mid p') = \sum_{1 \leq j \leq k} w_j' \cdot P_j(e \mid p'). 
\end{aligned}
\end{equation}

To explore the space of possible transformations and generate valid password candidates, 
we employ a beam search algorithm with beam width \( B \) 
as described in Algorithm~\ref{alg:mope_online}.

\begin{algorithm}
\footnotesize
    \caption{Online \name Password Generation}
    \label{alg:mope_online}
    \begin{algorithmic}[1]
    \STATE \textbf{Input:} Leaked password \(p'\), expert models \(\{\mathcal{M}_j\}_{j=1}^k\) in \name, beam width \(B\), candidates number $K$
    \STATE \textbf{Output:} Top-$K$ candidates \(\hat{p} = T(p', e)\) with scores \(P_{\text{\name}}(\hat{p})\)
    Calculate the weights \(\{w_j'\} = g(p')\)
    \FOR{each expert \(j\) with \(w_j' > 0\)}
        \STATE Perform beam search on \(P_j(e \mid p')\) using edit operations \(e_t \in \mathcal{O} = \{\texttt{del}, \texttt{ins}, \texttt{rep}, \texttt{end}\}\): \\
        \hspace{1em} Initialize beam \(B^{(0)} = \{(\varepsilon, 1)\}\), where \(\varepsilon\) is the empty edit sequence. \\
        \hspace{1em} For \(t = 1\) to max step, expand each \((e, P(e)) \in B^{(t-1)}\) with new operation \(e_t\): \\
        \hspace{2em} compute \(P_j(e') = P_j(e_t \mid p', e) \cdot P(e)\), where \(e' = (e, e_t)\). \\
        \hspace{1em} Keep top-\(B\) sequences to form \(B^{(t)}\), stop when all paths meet \texttt{end}. \\
        \hspace{1em} Let \(\mathcal{C}_j = \{(\hat{p}, P_j(\hat{p} \mid p'))\}\), where \(\hat{p} = T(p', e)\).
        
        \FOR{each \((\hat{p}, P_j(\hat{p} \mid p')) \in \mathcal{C}_j\)}
            \STATE \(P_{\text{\name}}(\hat{p} \mid p') \mathrel{+}= w_j' \cdot P_j(\hat{p} \mid p')\)
        \ENDFOR
    \ENDFOR
    \STATE \textbf{return} Top-$K$ \(\hat{p}\) sorted by \(P_{\text{\name}}(\hat{p} \mid p')\)
    \end{algorithmic}
    \end{algorithm}

\subsection{Distilling Offline \name}
\label{subsec:distill}
In real-world applications, the Password Strength Meters (PSMs) typically require real-time feedback, demanding a small model and fast inference. To meet these requirements, we propose an offline \name distillation method that compresses the offline \name into a smaller, more efficient version using distillation techniques~\cite{hinton2015distilling}. 

To this end, we first load the offline \name as the teacher model. 
For each input prefix \( x \), we use \name to generate a teacher distribution \( t(x) \), 
which represents the probability distribution over all characters in \(\Sigma\).
The student model adopts the same architecture as the expert model pre-trained on the overall password dataset,  
outputting a student distribution \( s(x) \) corresponding to \( t(x) \). 
To ensure \( s(x) \) approximates both the teacher distribution \( t(x) \) 
and the true label of the prefix \(x\), 
we fine-tune the student model using a hybrid distillation loss function defined as: 
\begin{equation}\nonumber
\mathcal{L}_{\text{Dist}} = \alpha \cdot \mathcal{L}_{\text{soft}} + (1 - \alpha) \cdot \mathcal{L}_{\text{hard}}. 
\end{equation}
The soft loss \( \mathcal{L}_{\text{soft}} \), which measures the divergence between teacher and student outputs,
is formulated using KL divergence, measuring the discrepancy between \(t(x)\) and \(s(x)\) with a temperature parameter \( T \) for smoothing, 
which is defined as: 
\begin{equation}\nonumber
\mathcal{L}_{\text{soft}} = T^2 \cdot \text{KLDiv}\left( \text{softmax}\left( \frac{t(x)}{T} \right) \;\|\; \text{softmax}\left( \frac{s(x)}{T} \right) \right).
\end{equation}
The hard loss \( L_{\text{hard}} \), which measures the discrepancy between the student output and the ground-truth labels, 
is formulated using
cross-entropy, measuring the discrepancy between \(s(x)\) and the true next character \( y \) of the prefix \(x\), 
which is defined as: 
\begin{equation}\nonumber
\mathcal{L}_{\text{hard}} = -\log s(x)_y, 
\end{equation}
where \( s(x)_y \) is the probability assigned by the student model to the true label, 
and \( \alpha \) is a parameter to balance soft and hard losses. To optimize the student model, we minimize the hybrid loss \( \mathcal{L}_{\text{Dist}} \) over the training dataset. 
The parameters \( \theta \) of the student model can be updated with learning rate \( \eta \) using gradient descent:
\begin{equation}\nonumber
\theta \leftarrow \theta - \eta \cdot \nabla_\theta \mathcal{L}_{\text{Dist}}. 
\end{equation}
After distillation, we obtain a single model \(\mathcal{M}_{\text{Dist}}\) taught by offline \name's knowledge, 
which can independently perform inference based on given prefixes and complete offline password guessing tasks. 
To generate passwords, 
we only need to replace the \name part in Algorithm~\ref{alg:mope_offline} with \(\mathcal{M}_{\text{Dist}}\). 

%% file: evaluation.tex
\section{EVALUATION}
\label{sec:evaluation}

In this section, we aim to answer the questions below. 

\begin{itemize}[label=•, leftmargin=*, labelsep=1em, itemindent=0em]
    \item How effective \name is for offline guessing, and what is the contribution of the fine-tuned expert models and center-distance-based gate method to the overall performance? (Section~\ref{subsec:offline_result})
    \item How effective \name is for online guessing, and what is the impact of the fine-tuned expert models and center-distance-based gate method on the overall performance? (Section~\ref{subsec:online_result})
    \item Why not use existing neural network-based techniques in \name's password clustering and gate method? (Section~\ref{sec:additional-evaluation})
\end{itemize}

\subsection{Experimental Settings}

\noindent\textbf{Implementation Details of Offline \name}. 
We use an LSTM layer with 256 hidden dimensions, a 64-dimensional embedding layer, 
and a fully connected layer with 128 dimensions as the architecture of our expert model. 
The number of expert models is consistent with the number of clusters from the training dataset. 
We pre-train each expert model on the training dataset for 30 epochs 
and fine-tune on the clusters respectively for 1 epoch using the Adam optimizer~\cite{DBLP:journals/corr/KingmaB14:adam}
with a batch size of 128, learning rate of 1e-3, dropout ratio of 0.15 
and a gate activation threshold $\beta$ of 10. 
It took us approximately 20 hours to finish clustering, 
training and fine-tuning on a dataset of 100,000 samples. 

For the distilled offline \name, 
the student model's architecture is the same as 
the pre-trained expert model in \name. 
Using offline \name as the teacher model, 
we randomly sample passwords from the dataset to distill \name to the student model 
using Adam optimizer, with $\alpha=0.7$, $T=2$, a learning rate of 1e-5, and a batch size of 128. 
For the \texttt{Rockyou} dataset, 20,000 passwords are sampled to distill, 
while for \texttt{CSDN} and \texttt{178} we randomly sample 30,000 passwords each. 
Distillation took approximately 10 hours per model.

\noindent\textbf{Implementation Details of Online \name}.
For online \name, we use a 3-layer LSTM encoder-decoder 
with 256-dimensional hidden and embedding sizes as the architecture of our expert model. 
The number of expert models is also consistent with the number of clusters from the training dataset. 
The expert model is pre-trained using the Adam optimizer 
with a batch size of 128 and dropout rate of 0.4 for 25 epochs, 
and is subsequently fine-tuned to specialize the expert models.
Fine-tuning is performed for 10 epochs, with a threshold \(\beta=2.5\) and a beam width \(B=150\) for inference. 
It took us approximately 10 hours to finish clustering, 
training and fine-tuning on a dataset of the order of 100,000 samples.

\noindent\textbf{Experiment Environment}. 
All experiments are conducted on a server running Ubuntu 20.04.2 LTS, 
equipped with a 40-core Intel Xeon Silver 4210 CPU operating at 2.20 GHz and 125 GB of RAM. 
We use an NVIDIA GeForce RTX 2080 Ti graphics card to accelerate the model's training 
and inference process.

\begin{figure*}[t]
  \centering
  \subfloat[\texttt{CSDN} $\rightarrow$ \texttt{Cit0day}]{
    \includegraphics[width=0.31\textwidth]{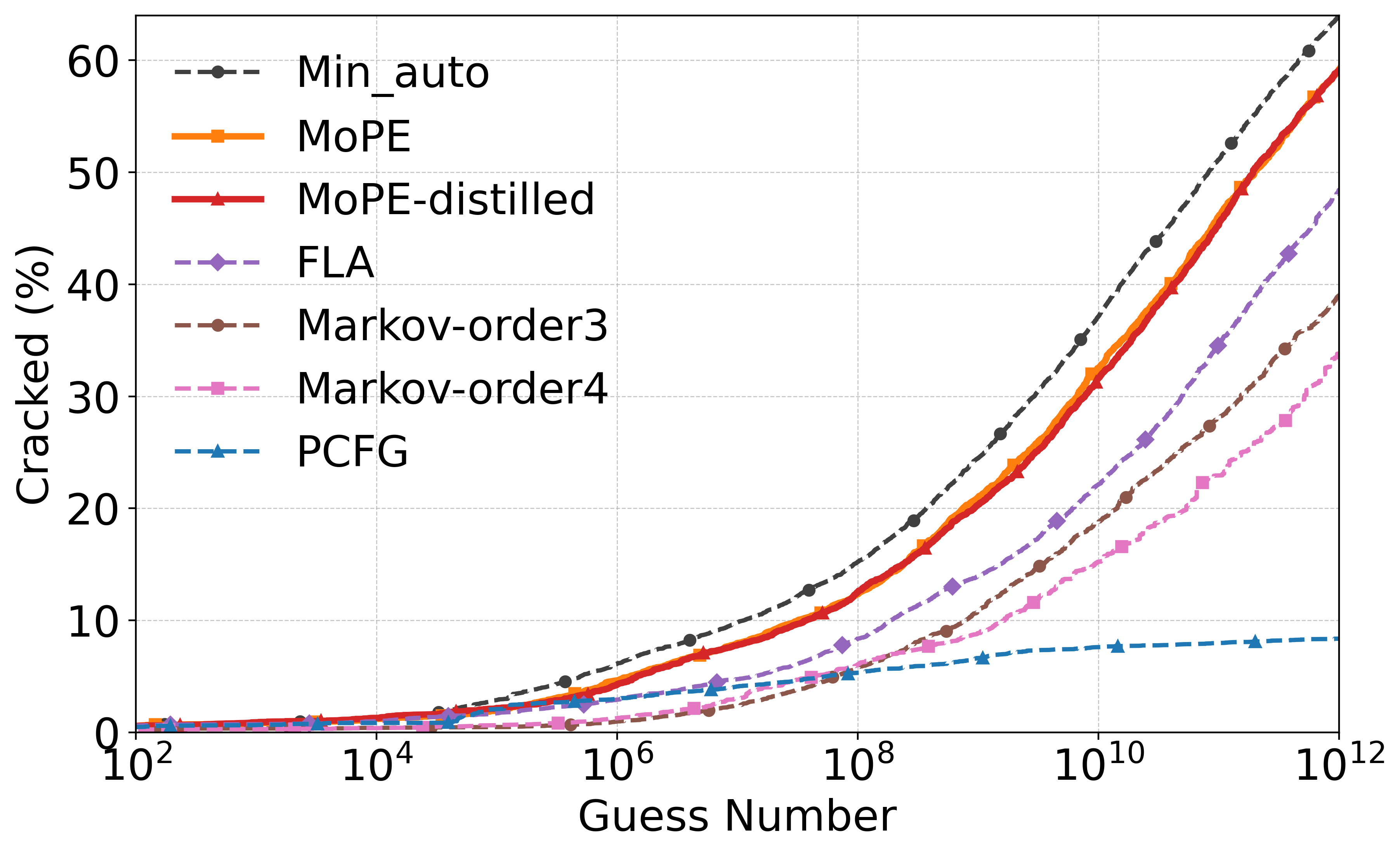}
  }
  \subfloat[\texttt{CSDN} $\rightarrow$ \texttt{Rockyou2024}]{
    \includegraphics[width=0.31\textwidth]{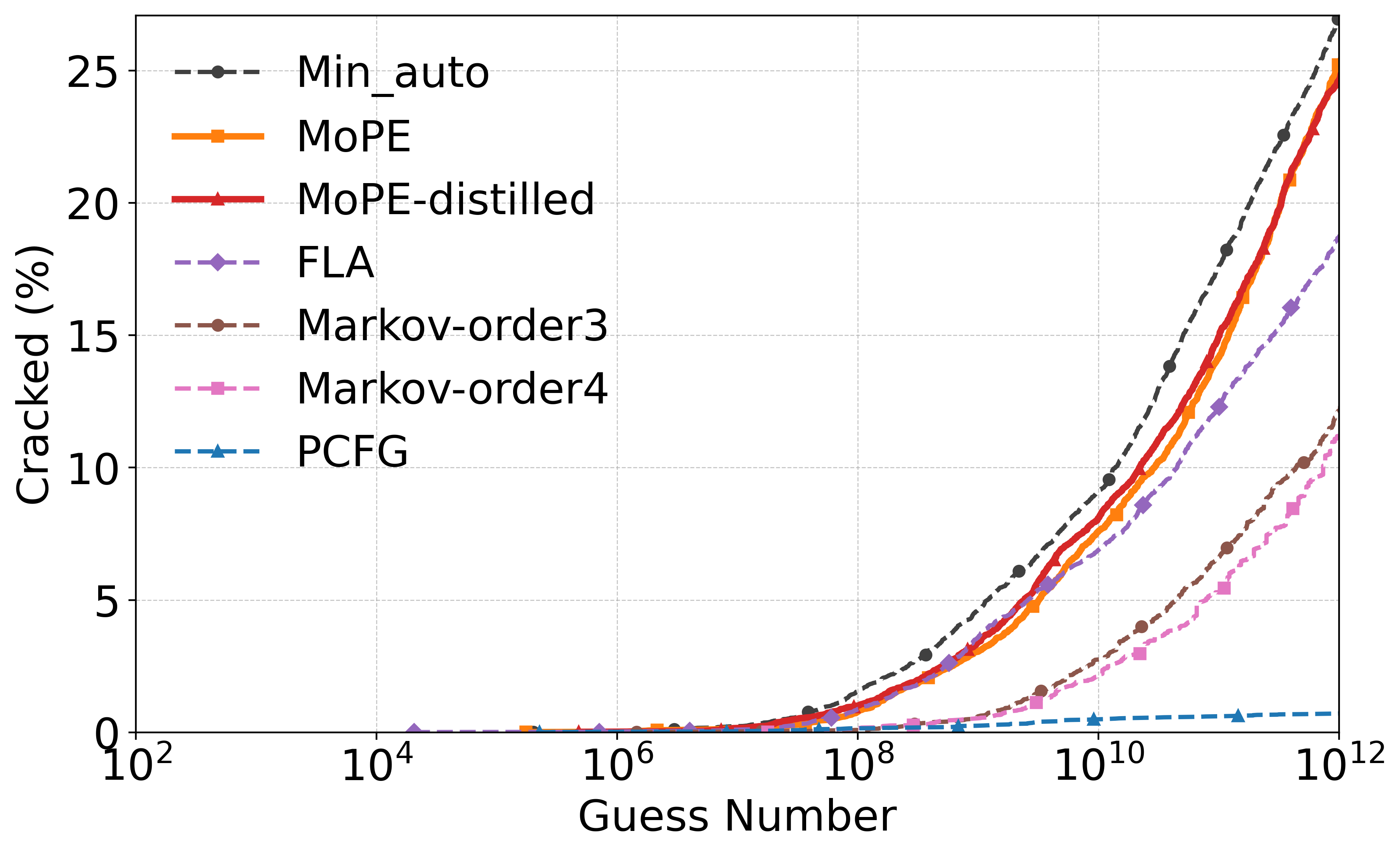}
  }
  \subfloat[\texttt{CSDN} $\rightarrow$ \texttt{Taobao}]{
    \includegraphics[width=0.31\textwidth]{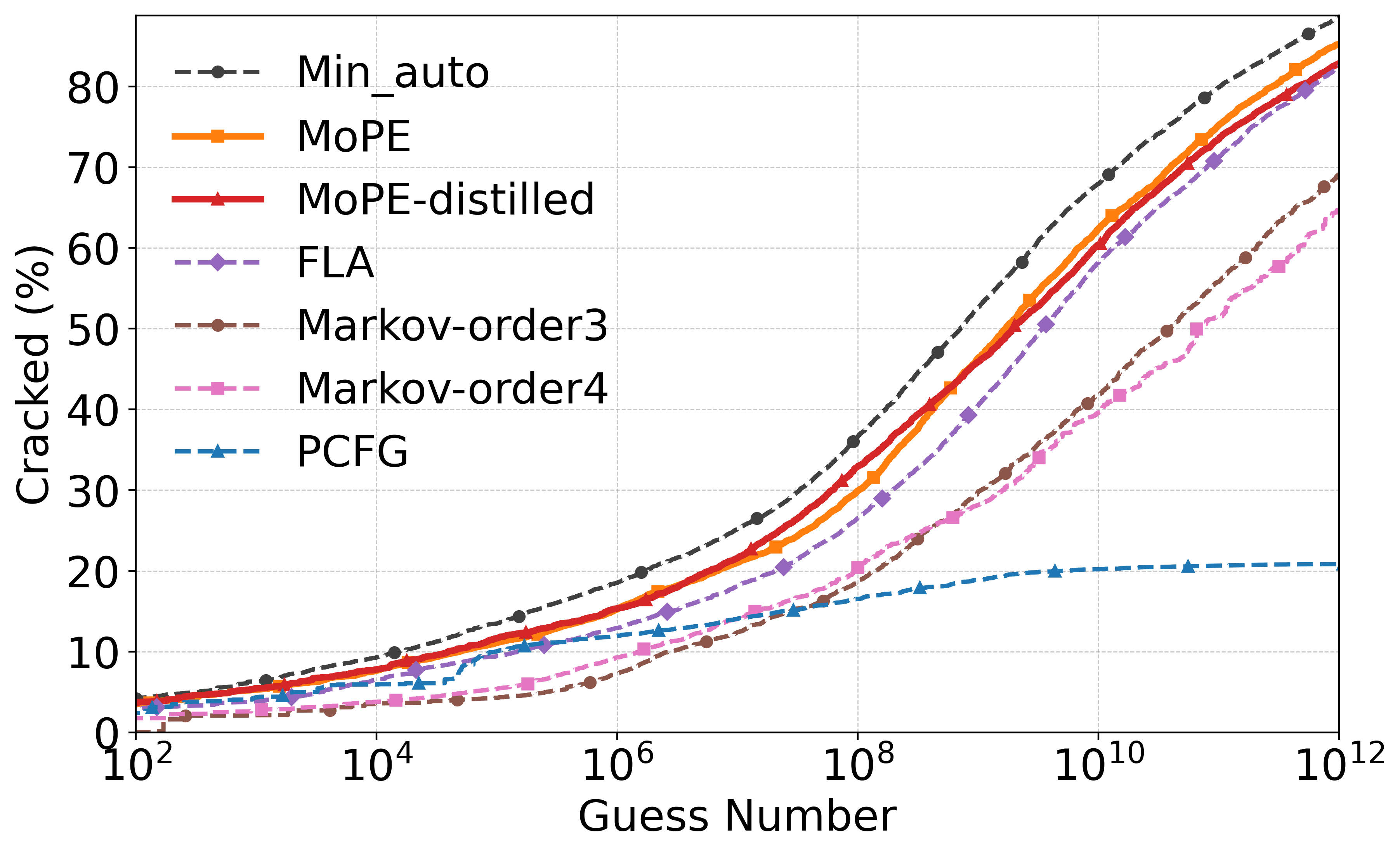}
  }

  \vspace{0.7em}

  \subfloat[\texttt{178} $\rightarrow$ \texttt{Cit0day}]{
    \includegraphics[width=0.31\textwidth]{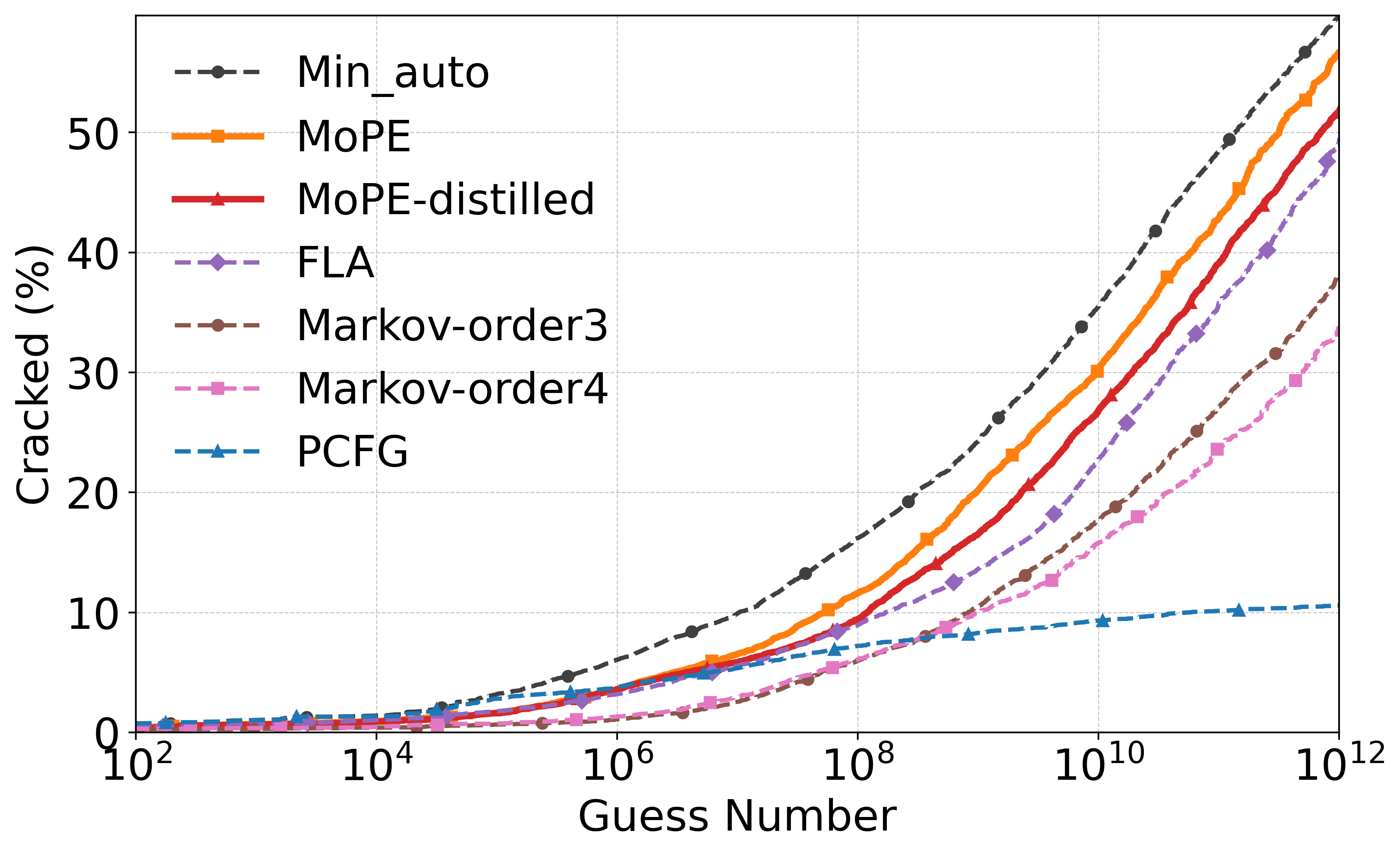}
  }
  \subfloat[\texttt{178} $\rightarrow$ \texttt{Rockyou2024}]{
    \includegraphics[width=0.31\textwidth]{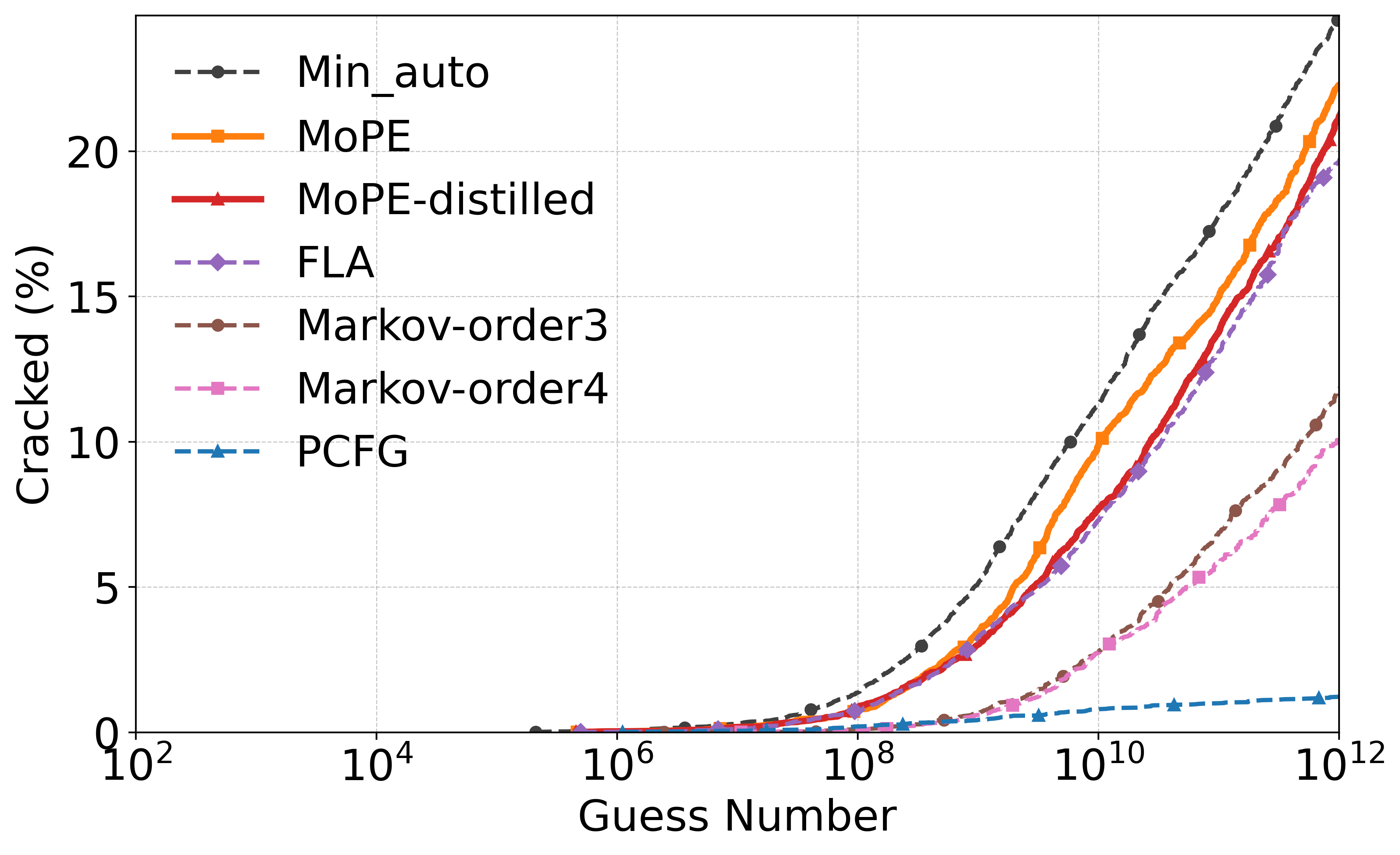}
  }
  \subfloat[\texttt{178} $\rightarrow$ \texttt{Taobao}]{
    \includegraphics[width=0.31\textwidth]{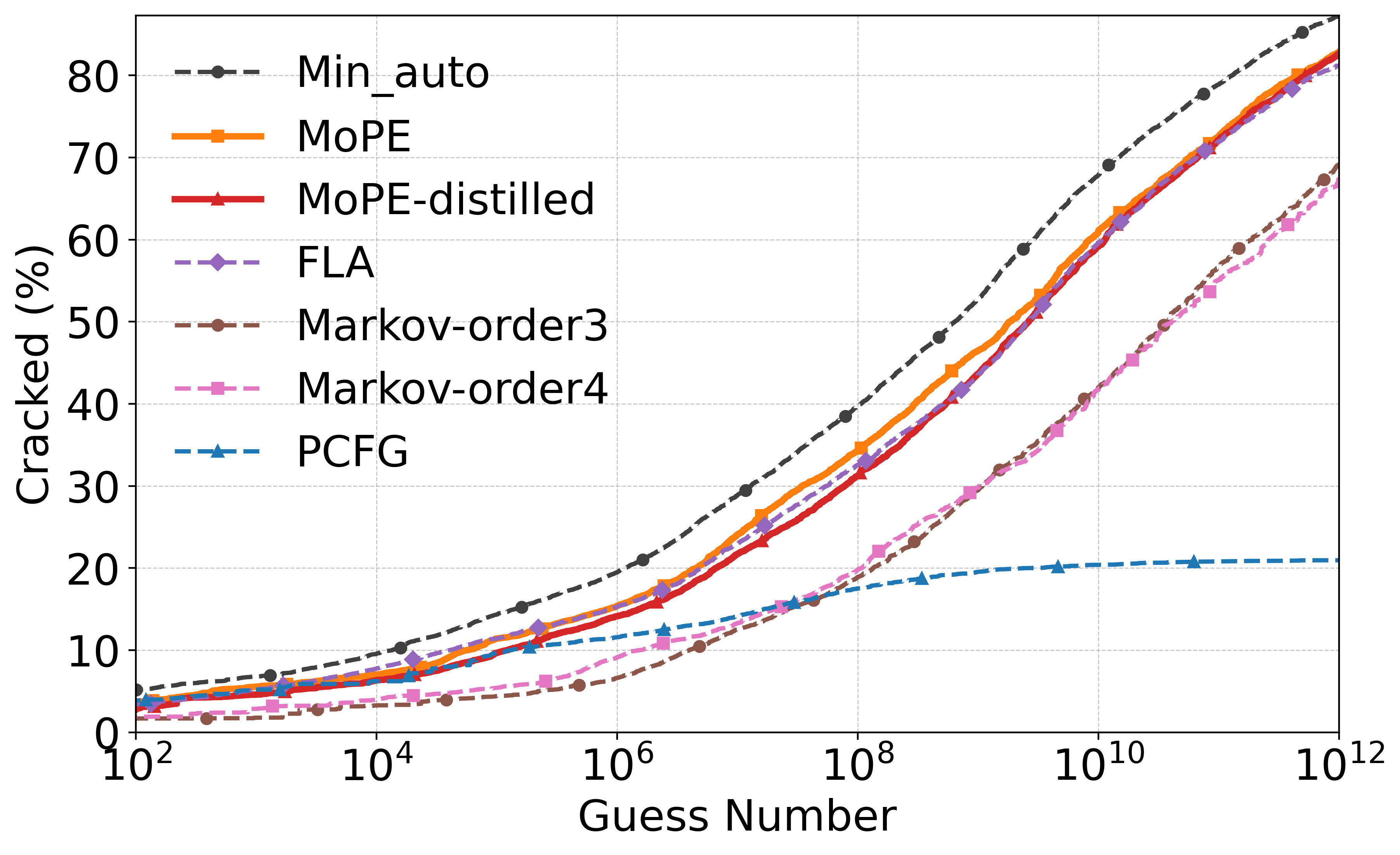}
  }

  \vspace{0.7em}

  \subfloat[\texttt{Rockyou} $\rightarrow$ \texttt{Cit0day}]{
    \includegraphics[width=0.31\textwidth]{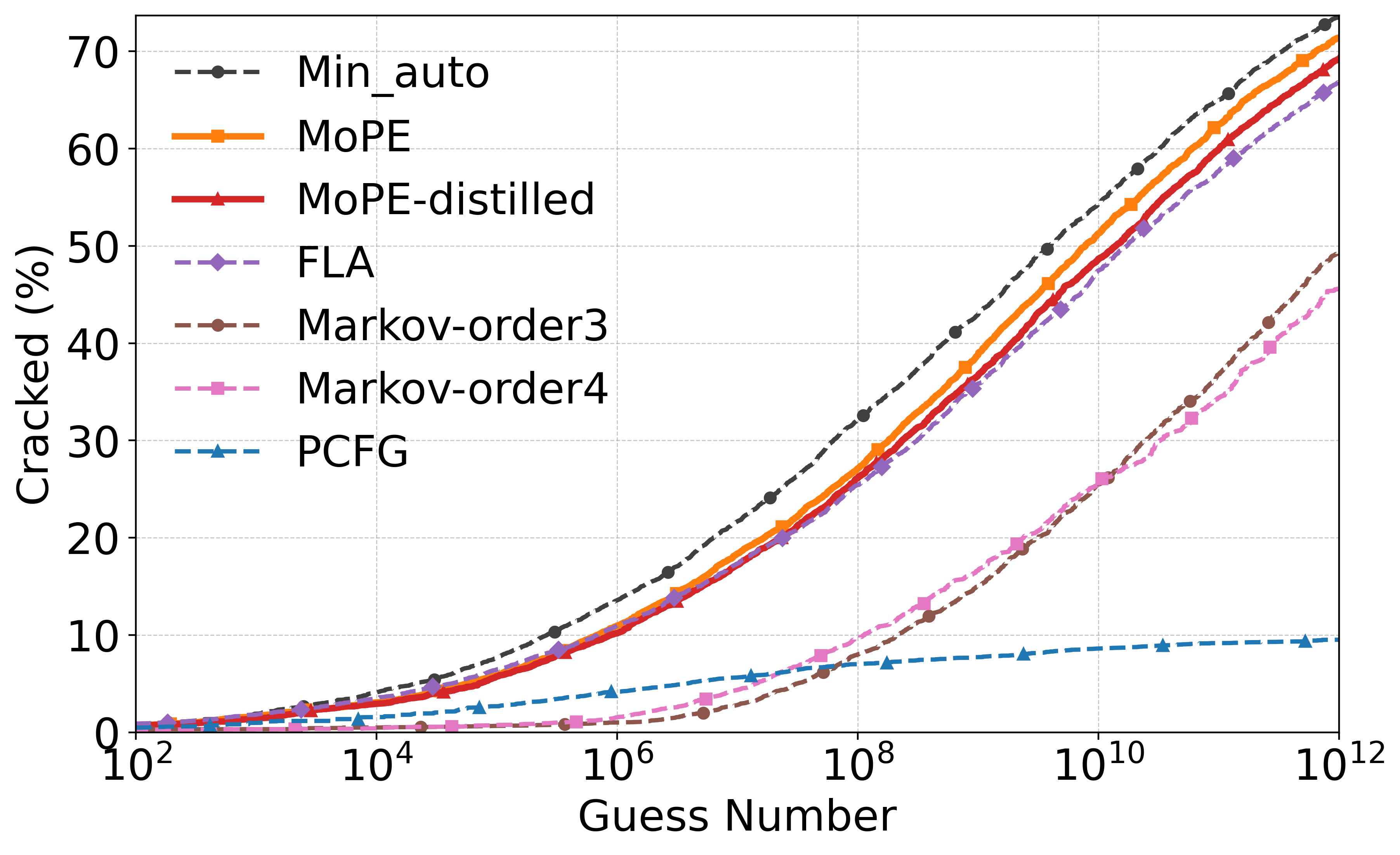}
  }
  \subfloat[\texttt{Rockyou} $\rightarrow$ \texttt{Rockyou2024}]{
    \includegraphics[width=0.31\textwidth]{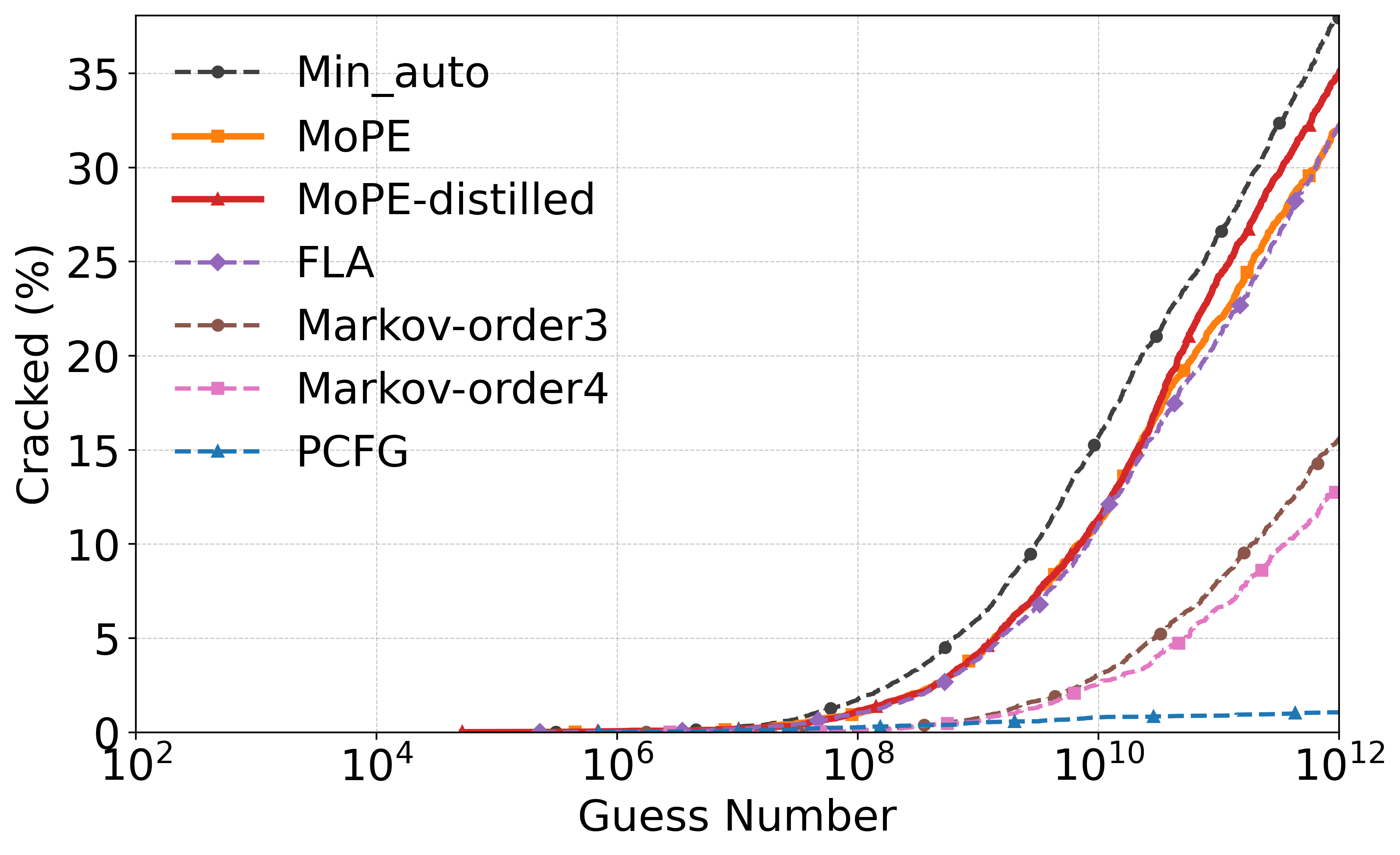}
  }
  \subfloat[\texttt{Rockyou} $\rightarrow$ \texttt{Taobao}]{
    \includegraphics[width=0.31\textwidth]{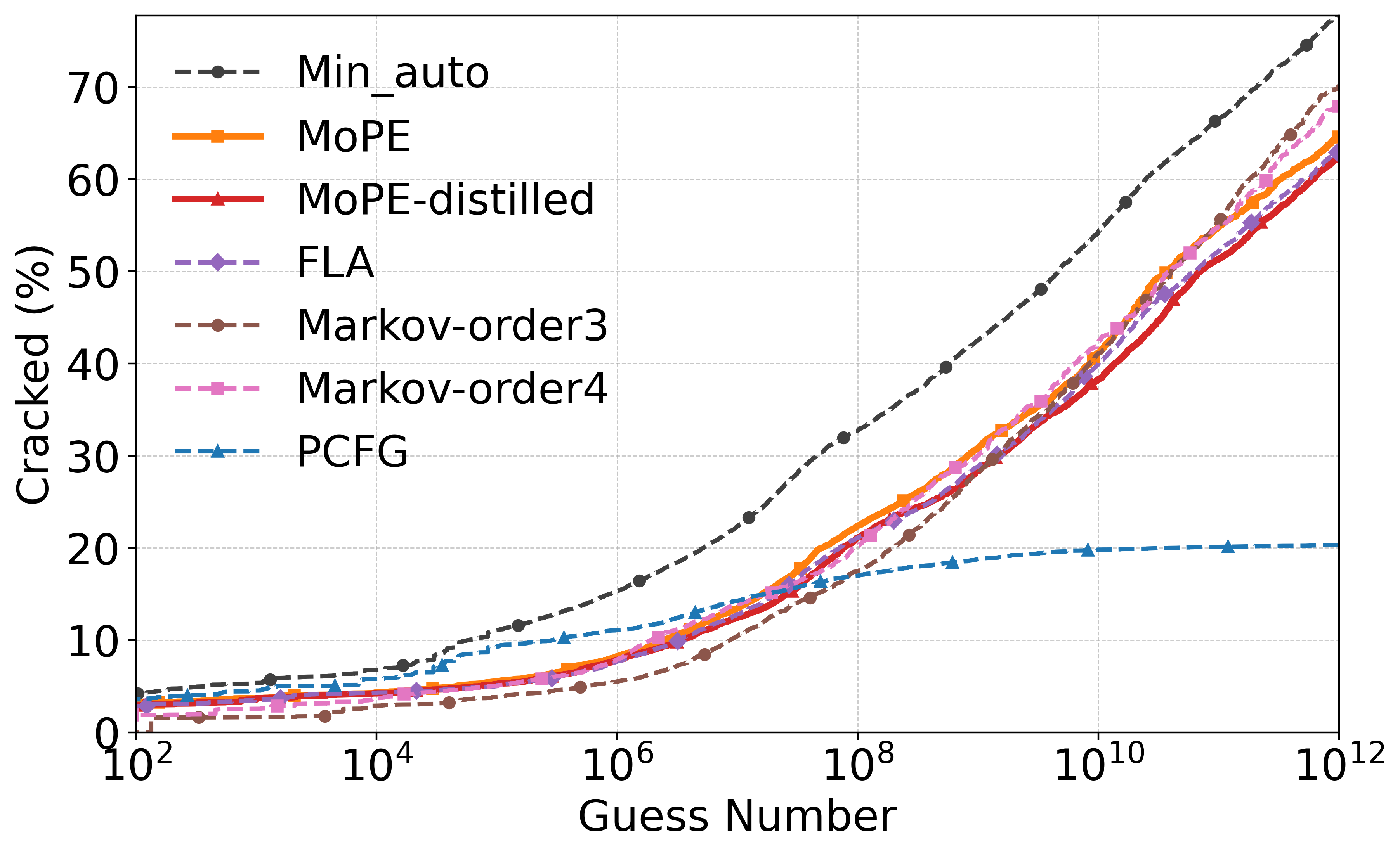}
  }

  \caption{
  Comparison results between \name and other SOTA methods (i.e., FLA, Markov, PCFG) in offline password guessing scenarios.
  In most cases, \textbf{\name most closely approaches the upper bound \textit{Min\_auto}, outperforming existing SOTAs}. 
  }
  \label{fig:offline_comparisons}
\end{figure*}

\subsection{Offline Password Guessing}
\label{subsec:offline_result}

\noindent\textbf{Evaluation Metrics}. 
We randomly sample 100,000 passwords from the offline password dataset as the training set, 
and 10,000 as the test set without overlap to avoid the potential evaluation bias. 
We simulate the offline scenario where the attacker generates up to $10^{12}$ password candidates, 
and compare the proportion of successful guesses among the test set as the cracking rate (\(\frac{N_{\text{cracked}}}{N}\)), 
using Monte-Carlo algorithm~\cite{DellAmico:Monter_Carlo} to calculate the cracking rate.

\noindent\textbf{Evaluation Baselines}. 
We choose mainstream offline password guessing methods as the baseline models with settings below.
Here are the settings:

\begin{itemize}[fullwidth,itemindent=0em]
    \item \textbf{PCFG~\cite{WeirPCFG}}: 
    The PCFG code we implement is based on the open source code~\cite{pcfg_cracker} on GitHub. 
    \item \textbf{Markov~\cite{NarayananMarkov}}: 
    We use 3-gram and 4-gram Markov models with Laplace smoothing for unseen n-grams.  
    \item \textbf{FLA~\cite{William:LSTM}}: 
    We implement FLA using PyTorch, with an embedding layer of 64 dimensions, 
    an LSTM layer with 256 hidden dimensions, 
    and a fully connected layer with 128 dimensions. 
    The model is trained for 30 epochs using Adam optimizer 
    with a batch size of 128, learning rate of 1e-3, and dropout ratio of 0.15. 
\end{itemize}

We also employ the \textit{Min\_auto}~\cite{cracking:usenix15} strategy, which represents an idealized guessing approach: a password is considered to be guessed using all these guessing models, serving as an upper bound of current password guessing performance. 

\noindent\textbf{Accuracy Results}. 
After performing clustering on passwords from the training set,  
we obtain 50, 35 and 55 clusters on \texttt{CSDN}, \texttt{178} and \texttt{Rockyou} respectively for \name, 
along with 50, 35 and 55 corresponding expert models. 
To comprehensively validate the effectiveness of our \name framework, 
we conduct experiments~\cite{DellAmico:Monter_Carlo} under a variety of cross-domain scenarios 
containing different languages and leakage periods. 
As shown in Figure~\ref{fig:offline_comparisons}, 
\name most closely approaches the \textit{Min\_auto} and outperforms existing SOTAs in most cases.
\name outperforms the SOTA model FLA in most scenarios, 
with improvements ranging from 0.52\% to 38.80\% (avg. 11.17\%).
The distilled \name, using the same architecture as FLA, 
surpasses FLA in most cases and, in some scenarios, 
even approaches the performance of the full \name. 
Especially in cross-lingual scenarios (e.g., subfigures a, b, d, e, and i), 
\name tends to outperform FLA more significantly. 
Moreover, in the evaluation on the newly collected \texttt{Rockyou2024} dataset, 
where passwords tend to follow rare structural patterns, 
\name achieves a substantial performance gain over all baselines, 
highlighting its remarkable generalization ability.

\noindent\textbf{Accuracy Analysis}. 
Different password datasets often exhibit distinct structural pattern distributions, leading the model to be biased toward frequent structural patterns of the training dataset. 
\name mitigates this bias through its structure-aware expert model design 
and high-accuracy gate method. 
Instead of relying on a single model, \name dynamically activates 
the relevant experts based on the intermediate prefix, 
including those trained on minority patterns in the training data. 
This design proves especially effective in cross-distribution scenarios (e.g., cross-lingual or across time periods). For example,
when transferring from Chinese datasets (\texttt{CSDN}, \texttt{178}) 
to English-based datasets (\texttt{Cit0day}, \texttt{Rockyou2024}) (Figure~\ref{fig:offline_comparisons}a, b, d, e), 
\name shows a significantly larger improvement over FLA compared to same-language transfers, 
indicating its ability to generalize beyond the dominant structures in training. 
Similarly, when the training data is outdated and fails to cover the features of the passwords well (as in the case of \texttt{Rockyou} $\longrightarrow$ \texttt{Rockyou2024}), traditional FLA struggles to crack the passwords using the learned features. 
Our \name model still outperforms FLA due to its more robust adaptation to the diverse password patterns in the test set. 
When the training and testing distributions are more aligned (e.g., Chinese-to-Chinese or English-to-English transfers), the gap between \name and FLA narrows, while \name still consistently outperforms FLA. 

\begin{figure}[h]
\setlength{\abovecaptionskip}{0pt}
\setlength{\belowcaptionskip}{0pt} 
    \centering
    \includegraphics[width=0.7\columnwidth]{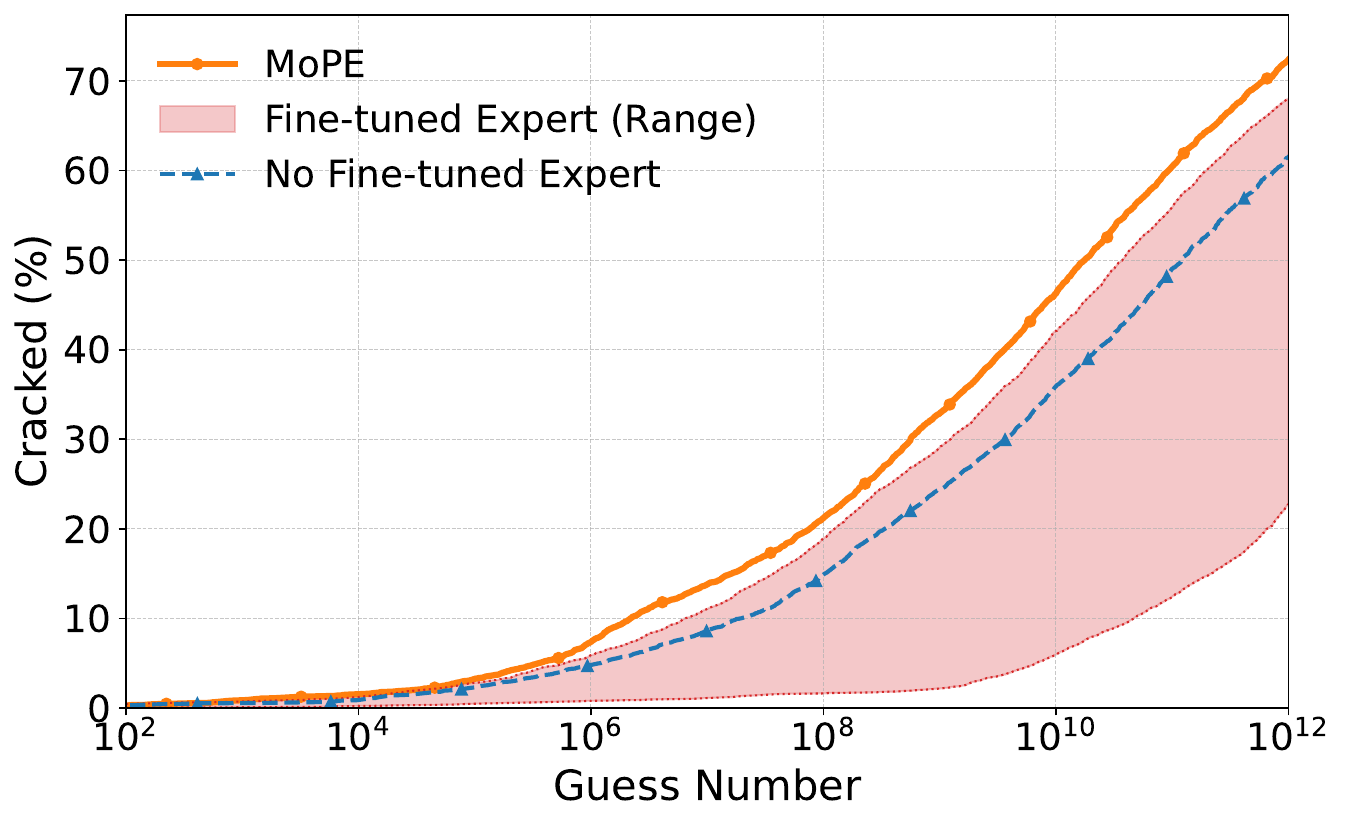}
    \caption{ 
    Ablation study of several experiment variants in offline \name.
    }
    \label{fig:10}
\end{figure}

\noindent\textbf{Ablation Study.} 
To understand the impact of expert specialization and gate design,  
we compare the cracking performance of several experiment variants: 
1) the no fine-tuned expert model (i.e., the expert model without fine-tuning), 
2) all fine-tuned expert models (i.e., each fine-tuned expert model is applied to produce one guessing performance curve). 
and 3) full \name, where the fine-tuned models are invoked dynamically by our center-distance-based gate method to produce the final guessing performance curve.  
We use \texttt{CSDN} as the training set (with 50 clusters generated) and \texttt{Neopets} as the test set. 
The results are shown in Figure~\ref{fig:10}, where the red shaded area represents the performance range (minimum to maximum) across all fine-tuned expert models.

\textit{(1) Fine-tuning on expert models.}
Several fine-tuned models outperform the no fine-tuned expert model, 
indicating that they are better aligned with the distribution of the test set after being fine-tuned on specific clusters. 
This variability arises because each expert model is fine-tuned to specialize in a particular structural pattern, 
thus leading to large performance range among fine-tuned experts.  

\textit{(2) Impact of center-distance-based gate method.} 
Our proposed center-distance-based gate achieves significantly better results in \name, 
outperforming the no fine-tuned baseline by nearly 20\% and surpassing the upper bound among all fine-tuned expert models.  
This indicates that \name's advantage 
does not stem from an isolated expert but a dynamic and context-aware expert selection method, highlighting the effectiveness and robustness of our gate method in leveraging structural patterns adaptively.

\subsection{Online Password Guessing}
\label{subsec:online_result}

\noindent\textbf{Experimental Configurations.} 
Our configurations are grounded in the assumption that attackers can obtain one leaked password of a specific user, 
and use it to crack another password of the same user. 
We randomly sample password pairs with two passwords from the same email 
and use one of them as the source password to crack the other. 
We extract source-target password pairs with an edit distance no greater than 4 
from the \texttt{Collection\#1} dataset, and randomly sample 100,000 pairs as the training dataset. 
For evaluation, we randomly sample 10,000 source-target password pairs 
from the \texttt{4iQ} and \texttt{Collection\#1} datasets, respectively, with each pair corresponding to a unique user account.
The proportion of emails from the training set 
appearing in the test set is 0.02\% and 0\%
in \texttt{4iQ} and \texttt{Collection\#1} respectively, leaving the overlap impact to be negligible.

\noindent\textbf{Evaluation Metrics}. 
During testing, given the source password as input, 
the model generates a list of password candidates 
under different guessing budgets (10, 100, and 1,000). 
If the target password appears in the generated candidate list, 
we consider the corresponding account successfully cracked. 
The final crack rate is calculated as \(\frac{N_{\text{cracked}}}{N_{\text{account}}}\).

\noindent\textbf{Evaluation Settings}.
We select existing open models, namely TarGuess-II~\cite{DBLP:conf/ccs/WangZWYH16}, 
Pass2path~\cite{DBLP:conf/sp/PalD0R19:similarity}, and PassBERT~\cite{xu-real-world-guessing}, 
to verify the effectiveness of our online \name. 
Here are the settings:
\begin{itemize}[fullwidth,itemindent=0em] 
    \item \textbf{TarGuess-II}. 
    We use the whole \texttt{RockYou} dataset (introduced in Section~\ref{subsec:password_datasets}) 
    as the source of popular passwords, 
    and set the popularity list size to 10,000 with the similarity threshold of 0.5.

    \item \textbf{Pass2path}. 
    The Pass2path model~\cite{DBLP:conf/sp/PalD0R19:similarity} adopts a 3-layer LSTM encoder-decoder structure with 256-dimensional hidden and embedding sizes. 
    The model is trained using the Adam optimizer with a learning rate of 0.01, a batch size of 128, a dropout rate of 0.4, 
    and is trained for 25 epochs. 

    \item \textbf{PassBERT}. 
    We adopt the medium configuration of PassBERT~\cite{xu-real-world-guessing}, 
    using a 4-layer BERT model. 
    The model is trained using the Adam optimizer with a batch size of 128 and a learning rate of 1e-5. 
    Beam search is applied with a beam width of 150.

\end{itemize}

\begin{table}[htbp]
\footnotesize
\setlength{\abovecaptionskip}{0pt}
\setlength{\belowcaptionskip}{0pt}
    \centering
    \caption{Cracking rates in online guessings.}
    \begin{tabular}{c|ccc|ccc}
    \toprule
    \textbf{Attack model} & \multicolumn{3}{c|}{ \texttt{4iQ} (\%)} & \multicolumn{3}{c}{\texttt{Collection\#1} (\%)} \\
     & 10 & 100 & 1,000 & 10 & 100 & 1,000 \\
    \midrule
    \name & \cellcolor{gray!30}10.72 & \cellcolor{gray!75}15.00 & \cellcolor{gray!75}17.45 & \cellcolor{gray!30}8.64 & \cellcolor{gray!75}14.16 & \cellcolor{gray!75}16.93 \\ 
    PassBERT & \cellcolor{gray!75}11.15 & \cellcolor{gray!30}14.22 & \cellcolor{gray!45}16.79 & \cellcolor{gray!75}9.05 & \cellcolor{gray!30}12.96 & \cellcolor{gray!45}16.22 \\
    Pass2path & \cellcolor{gray!45}10.94 & \cellcolor{gray!45}14.67 & \cellcolor{gray!30}16.05 & \cellcolor{gray!45}8.74 & \cellcolor{gray!45}14.09 & \cellcolor{gray!30}15.65\\
    TarGuess-II & 8.62 & 10.97 & 13.30 & 4.88 & 8.37 & 10.88 \\
    \bottomrule
    \end{tabular}
    \label{tab:online_result}
\end{table}

\begin{figure}[h]
    \centering
    \includegraphics[width=0.9\columnwidth]{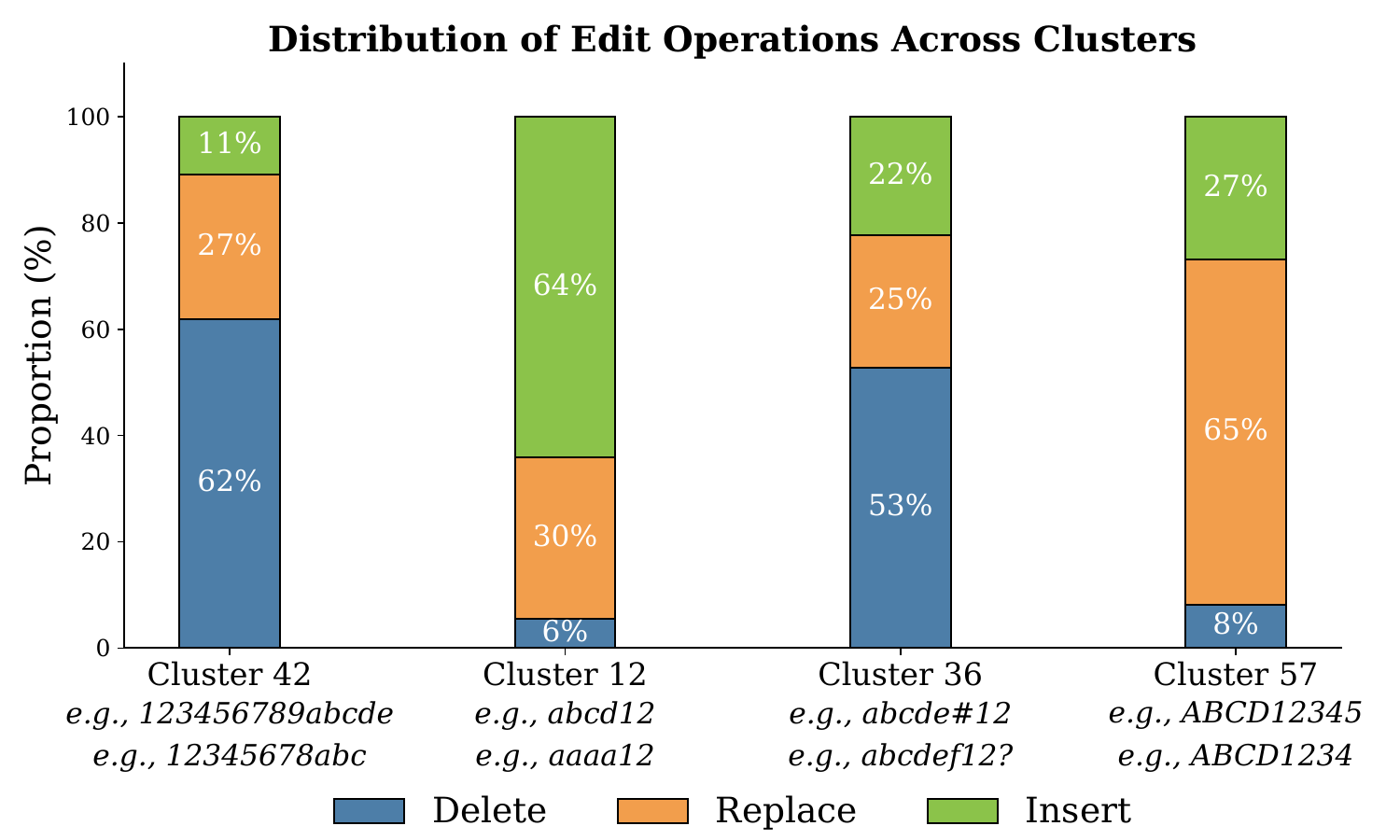}
    \caption{
        Distribution of edit operations (Delete, Replace, Insert) across 4 randomly sampled clusters, indicating that \textbf{different password structures favor specific transformation rules}. 
    }
    \label{fig:8}
\end{figure}

\noindent\textbf{Evaluation Results}. The performance is reported in Table~\ref{tab:online_result}.  
Our \name achieves a higher cracking rate within 100 guesses and 1,000 guesses,  
outperforming the SOTA PassBERT by 5.49\%$\sim$9.27\% (avg. 5.77\%).  
The relatively modest performance of \name within top-10 guesses compared to PassBERT, stems from \name's tendency to introduce a broader range of relatively complex edit paths early in the guessing process, while the actual edit paths of some samples are prone to be simple within less budgets. 
In contrast, models like PassBERT are more likely to prioritize much simpler transformations—such as inserting a character or making no change at all (e.g., \{insert “1”\}, or~\{$\varnothing$\})—which are more effective in the earliest guesses, since a small number of samples in the dataset can be cracked with minimal changes. 
Still, \name outperforms PassBERT in both top-100 and top-1000 guesses with relatively more complex edit paths (e.g., \{insert “1”, insert “2”, insert “3”\}), indicating its strength in modeling more sophisticated transformation patterns.

\noindent\textbf{Analysis}. 
We obtain 65 distinct clusters for online \name after performing clustering on source passwords from the training set, meaning 65 corresponding expert models. 
We randomly sample four clusters and analyze the distribution of the basic edit operation types from source to target passwords in Figure~\ref{fig:8}, suggesting that each cluster has their own edit operation preferences. 
Passwords in cluster 42 have longer lengths,  
tending to involve more \texttt{del} operations, while shorter passwords in the others are more associated with \texttt{ins} operations.  
Passwords in cluster 57 have a higher proportion of uppercase characters, 
exhibiting a greater tendency for \texttt{rep} operations, compared to the others without uppercase characters. For example, password ``abcd12'' is largely modified by \texttt{insertion}, while ``123456789!'' favors \texttt{deletion}, thus enabling our online \name to offer proactive and structure-specific suggestions.

We reasonably attribute the effectiveness of online \name 
to its ability to model the nuanced transformation paths across structurally distinct password clusters. 
Certain structures exhibit a preference for specific transformation rules (shown in Figure~\ref{fig:8}).
Training models directly on the full dataset tends to bias them toward the dominant structural patterns,  
thereby limiting their effectiveness when dealing with less common password structures. 
Our online \name explicitly estimates the structural pattern of each source password  
and assigns higher weights to the corresponding expert models. 
This can prevent over-reliance on any single model 
and promotes diversity in transformation modeling. 
By tailoring the transformations to specific cluster patterns, 
online \name can make more accurate predictions, 
thereby improving overall cracking performance.


\begin{figure}[h]
\setlength{\abovecaptionskip}{0pt}
\setlength{\belowcaptionskip}{0pt} 
    \centering
    \includegraphics[width=0.7\columnwidth]{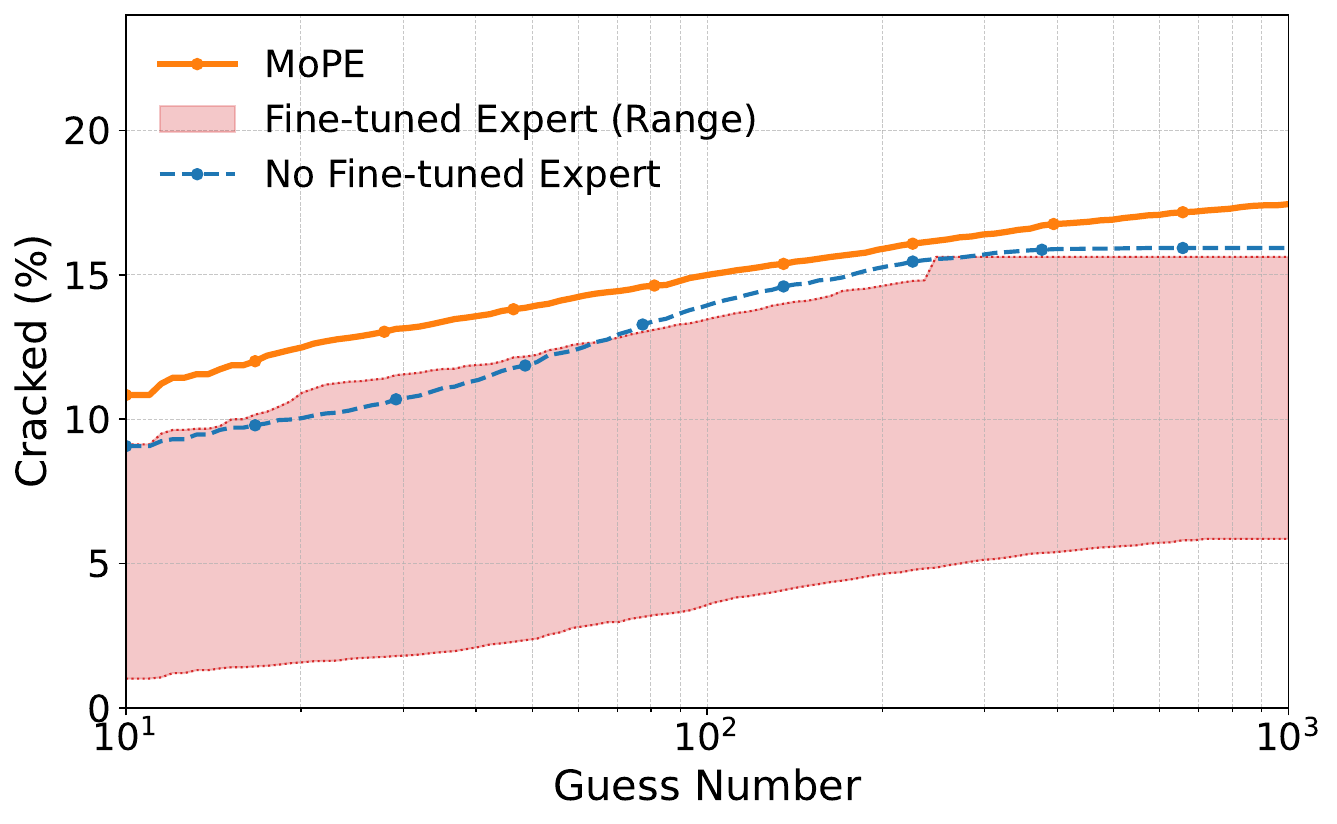}
    \caption{
      Ablation study of several experiment variants in online \name.
      }
    \label{fig:11}
\end{figure}

\noindent\textbf{Ablation Study.} 
We perform a similar ablation study in online guessing with three experimental variants: 
1) the no fine-tuned expert model, 
2) all fine-tuned expert models, 
and 3) \name that supports dynamically activating relevant fine-tuned expert models based on our center-distance-based gate method.  
We use \texttt{Collection\#1} as the training set (with 65 clusters generated) and \texttt{4iQ} as the test set. 
The results are shown in Figure~\ref{fig:11}.
The performance of fine-tuned expert models varies widely, as each model exhibits a distinct guessing curve.
The online dataset is a mixture of data from multiple platforms and regions, resulting in greater distributional complexity compared to the offline scenario. 
As a result, any single structural transformation pattern struggles to generalize well to the online test set, 
thus the upper bound performance of fine-tuned experts tends to be close to no fine-tuned expert model. Still, \name outperforms all the variations, 
highlighting the limitations of using fine-tuned expert models without structure-aware selection 
and our center-distance-based gate can instead enable better performance 
by adaptively selecting expert models based on the structural features of input contexts. 

By combining the ablation results from both offline and online scenarios, 
we arrive at the finding that 
\textbf{both expert specialization and center-distance gate-based expert selection are essential components of \name, 
enabling better adaptation to structural diversity and password modeling.}

\subsection{Comparison Study with Neural network-based Techniques}~\label{sec:additional-evaluation}
We compare the performance of neural network-based techniques 
for password clustering and gate methods. 

\noindent\textbf{Password Clustering: Structure-based vs. Neural Network-based.} 
We argue that neural network-derived (e.g., FLA~\cite{William:LSTM}) high-dimensional vectors
fail to generate differentiated expert models. 
To verify this claim, we compare the two methods' performance in silhouette score for assessing clustering quality. 
As shown in Table~\ref{tab:silhouette_cmp}, 
our structure-based clustering significantly outperforms the FLA-based method across various cluster numbers 
on both \texttt{CSDN} and \texttt{RockYou}. 
This confirms that 
our structure-based clustering provides better clustering quality, 
which is essential for enabling expert diversity and performance in \name's password modeling.

\begin{table}[ht]
\setlength{\abovecaptionskip}{0pt}
\setlength{\belowcaptionskip}{0pt}
\footnotesize
\centering
\renewcommand\tabcolsep{9.2pt}
\caption{Silhouette Scores of clustering: FLA-based vs. structure-based (Ours).}
\label{tab:silhouette_cmp}
\begin{tabular}{cccc}
\toprule
\textbf{Clusters} & \textbf{CSDN (FLA / Ours)} & \textbf{RockYou (FLA / Ours)} \\
\midrule
30 & 0.1337 / 0.6046 & 0.0230 / 0.5945 \\
35 & 0.1037 / 0.6672 & 0.0352 / 0.6195 \\
40 & 0.1148 / 0.6777 & 0.0283 / 0.6695 \\
45 & 0.1073 / 0.6837 & 0.0296 / 0.6943 \\
50 & 0.1025 / 0.7072 & 0.0232 / 0.6978 \\
\bottomrule
\end{tabular}
\end{table}

\begin{table}[ht]
\setlength{\abovecaptionskip}{0pt}
\setlength{\belowcaptionskip}{0pt}  
\centering
\caption{Gate inference efficiency on 10,000 samples. 
}
\label{tab:gate-efficiency}
\begin{tabular}{lccc}
\toprule
\textbf{Method} & \textbf{Total Time (ms)} & \textbf{Avg Time (ms)} & \textbf{Samples/s} \\
\midrule
NN (LSTM, CPU) & 17,573 & 17.57 & 56.90 \\
NN (LSTM, GPU) & 470 & 0.47 & 2,130 \\
\textbf{Ours (CPU)} & \textbf{44} & \textbf{0.04} & \textbf{22,920} \\
\bottomrule
\end{tabular}
\end{table}

\noindent\textbf{Gate Design: Center-distance-based vs. Neural Network-based.}
We compare the efficiency of two gate methods by replacing our lightweight center-distance-based gate 
with a recurrent neural network-based gate using LSTM~\cite{William:LSTM, DBLP:conf/sp/PalD0R19:similarity}. 
~\cite{William:LSTM, DBLP:conf/sp/PalD0R19:similarity}.
We implement an LSTM gate with 256 hidden dimensions and a 64-dimensional embedding layer, 
evaluating its performance against our method on 10,000 randomly sampled password samples.
We measure total runtime, average latency per sample, and throughput in terms of samples processed per second. 
The LSTM gate is tested on both CPU and CUDA-enabled GPU, 
while our center-distance gate is purely CPU-based. 
Results are shown in Table~\ref{tab:gate-efficiency}.
Our method achieves over \textbf{400$\times$ speedup} compared to the CPU-based neural gate 
and more than \textbf{10$\times$ faster} than its GPU-accelerated counterpart. 
This speed improvement is critical for \name, 
where the gate must select experts for every given password or prefix during inference, 
significantly boosting the overall efficiency of \name. 
Password guessing typically involves evaluating millions to billions of candidates in offline and online scenarios, making the computationally intensive neural network-based approaches infeasible. 
By enabling high-throughput expert selection with minimal overhead, 
our gate ensures that \name remains responsive and scalable in large-scale scenarios in real-world deployments. 

\section{PASSWORD STRENGTH METER}~\label{sec:psm}
A variety of PSMs~\cite{DBLP:conf/ndss/CastellucciaDP12:adaptive-markov, William:LSTM, DBLP:conf/sp/PalD0R19:similarity, DBLP:conf/esorics/PasquiniAB20:CPGmeter, zxcvbn, xu2021chunk, yang2024rankguess} 
have been proposed to 
evaluate password strength, enhancing users' awareness of potential guessing attacks. 
We use the distilled \name \(\mathcal{M}_{\text{Dist}}\) to implement \textbf{\name-PSM}, 
which requires only a single expert model, 
significantly reducing computational overhead and deployment complexity. Distilled \name achieves \textbf{20$\times$} lower query latency than the offline \name, maintaining comparable latency with FLA and producing fewer unsafe errors than FLA. 

\noindent\textbf{Comparison of Unsafe Errors.} 
We benchmark our \name-PSM against FLA-PSM~\cite{William:LSTM} with both models configured to have an equal number of parameters.  
To measure the precision of both PSMs, we adopt 
unsafe errors~\cite{William:LSTM}, which refer to passwords that can be easily guessed with fewer attempts but are incorrectly rated as strong by the PSM. 
This type of overestimation is particularly dangerous, 
as it may falsely reassure users and discourage them from updating weak passwords~\cite{DBLP:conf/chi/UrBSBCC16:perceptionsofpwd}. 
We classify password strength
into three levels~\cite{DBLP:journals/cacm/FlorencioHO16:guess-strength}: 
\textit{weak} (guesses $ \textless 10^6$), 
\textit{medium} ($10^6 \leq$ guesses $ \textless 10^{14}$), 
and \textit{strong} (guesses $\geq$ $10^{14}$).  
We randomly sample 100,000 passwords from \texttt{Neopets} 
and evaluate the unsafe error of both \name-PSM and FLA-PSM. 
Table~\ref{tab:psm_rank_heatmap} shows that \name-PSM significantly reduces the unsafe error compared to FLA-PSM. 
Notably, approximately 5\% of passwords are rated as \textit{medium} by FLA-PSM, 
but are actually \textit{weak}, 
highlighting \name-PSM's better precision in rating passwords.

\begin{table}[h!]
\setlength{\abovecaptionskip}{0pt}
\setlength{\belowcaptionskip}{0pt}
    \centering
    \caption{
\textit{Unsafe errors} between \name-PSM and FLA-PSM. 
}
    \label{tab:psm_rank_heatmap}
    \renewcommand{\arraystretch}{1.2}
    \resizebox{\columnwidth}{!}{%
    \begin{tabular}{c|ccccccc}
    \toprule
     & \multicolumn{7}{c}{\textbf{FLA-PSM}} \\
    \cmidrule(lr){2-8}
    \textbf{\name-PSM} 
    & $>10^0$ & $>10^3$ & $>10^6$ & $>10^9$ & $>10^{12}$ & $>10^{15}$ & $>10^{18}$ \\
    \midrule
    $>10^0$      & \cellcolor{green!30}426   & \cellcolor{red!35}254    & \cellcolor{red!15}24     & \cellcolor{red!5}1     & 0     & 0     & 0 \\
    $>10^3$      & \cellcolor{blue!10}136   & \cellcolor{green!45}2751  & \cellcolor{red!70}4339   & \cellcolor{red!40}547   & 0     & 0     & 0 \\
    $>10^6$      & 0     & \cellcolor{blue!20}1014  & \cellcolor{green!90}13764 & \cellcolor{red!100}11899 & \cellcolor{red!30}468  & \cellcolor{red!5}1    & 0 \\
    $>10^9$      & 0     & \cellcolor{blue!5}1     & \cellcolor{blue!20}1416  & \cellcolor{green!100}22448 & \cellcolor{red!90}10261 & \cellcolor{red!35}254  & 0 \\
    $>10^{12}$   & 0     & 0     & 0     & \cellcolor{blue!20}1548  & \cellcolor{green!60}13804 & \cellcolor{red!60}2683  & \cellcolor{red!10}13   \\
    $>10^{15}$   & 0     & 0     & 0     & \cellcolor{blue!5}1     & \cellcolor{blue!20}853   & \cellcolor{green!55}5521  & \cellcolor{red!25}756   \\
    $>10^{18}$   & 0     & 0     & 0    & 0     & 0     & \cellcolor{blue!15}577   & \cellcolor{green!50}2271  \\
    \bottomrule
    \end{tabular}%
    }
\vspace{1mm}
{\footnotesize
\noindent\textit{Note.} For example, 4,339 passwords are rated as cracked in $10^3 \sim 10^6$ guesses by \name-PSM, while requiring more than $10^6$ guesses according to FLA-PSM. This indicates that FLA-PSM regards these passwords as safer, whereas \name-PSM estimates them to be weaker.
}
\end{table}

\section{DISCUSSION}
\label{sec:discussion}

\noindent\textbf{Theoretical Proof of \name}. 
We empirically show that \name can achieve significantly superior guessing performance. 
To establish a solid foundation for \name, 
we provide a theoretical proof that \name always has lower errors in inference than the traditional single model, 
with the assumptions of high-quality clustering and accurate gate selection methods (detailed in Appendix~\ref{sec:proof}). 
Under the assumptions, the ground-truth password distribution can be viewed as a finite mixture of cluster-specific subdistributions, and a calibrated gate aligns \name’s weights with those mixture proportions approximately. By the joint convexity of KL divergence, the overall divergence between the true distribution and \name’s output is upper-bounded by the weighted average of the per-cluster divergences between the true and predicted distributions. 
Because high-quality clustering yields structurally patterned subdistributions, each expert, specialized on its corresponding subdistribution, achieves a lower approximation error on that subdistribution compared to a single global model trained over the entire distribution. The aggregation of these experts—guided by the gate—preserves this reduction in loss,
leading to improved overall modeling accuracy.

\noindent\textbf{Relevance and Distinctions with MoE Architectures in LLMs.} Our \name framework is inspired by MoE but differs significantly from its application in LLMs. 
In both LLMs and \name, a gate method is employed to dynamically select the appropriate expert models based on the input; 
however, while MoE in LLMs addresses task diversity and semantic complexities (e.g., distinguishing between different types of language tasks) to improve performance without unnecessary computation,  
\name focuses on the structural diversity of passwords to improve overall password guessing performance. 
Our method applies MoE-inspired ideas to tackle the unique challenge of password guessing, specifically optimizing performance based on password structural patterns.

\noindent\textbf{Takeaways of \name}. 
With \name, we demonstrate that passwords, as diverse textual data, 
can be processed separately based on their structural patterns 
to improve the accuracy of password guessing modeling. 
The structure-aware modeling framework allows \name to more effectively capture the underlying patterns in password distributions, 
enhancing the generalization capability of password modeling in password structures. Furthermore, \name-PSM can reduce false security errors by more accurately assessing password strength with millisecond-level latency,
improving users' password security in real-time applications.

\section{CONCLUSION}
\label{sec:conclusion}
In this paper, we propose the \name framework, 
a novel approach for improving password guessing accuracy 
by leveraging the structural patterns within passwords. 
By dividing passwords into distinct clusters and fine-tuning expert models for each cluster, 
we achieve more precise password modeling. 
Additionally, we introduce a lightweight gate method to optimize computational efficiency 
without compromising accuracy. 
Our comprehensive evaluation demonstrates the effectiveness of the \name framework 
in both offline and online password guessing scenarios, achieving up to 38.8\% and 9.27\% improvement in cracking rate, respectively. 
We introduce a new Password Strength Meter (PSM) based on the distilled \name, 
offering users a more accurate way to assess password strength 
and enhance users' risk awareness. 

\clearpage
\section{Ethics Considerations}
This research aims to enhance password modeling and improve security 
using publicly accessible datasets. 
We strictly follow established ethical guidelines, with particular emphasis on the 
protection of user privacy. All datasets used in this study are 
handled in a manner that avoids storage, redistribution, leakage or 
exploitation beyond what is necessary for research purposes. 
Furthermore, no personally identifiable information is 
collected, retained, or disclosed throughout the study, ensuring 
compliance with responsible data usage and privacy standards.

We choose to open-source a partial key implementation of \name framework, 
following a common practice in prior work~\cite{William:LSTM, DBLP:conf/sp/PalD0R19:similarity, xu-real-world-guessing} to foster iterative 
development within the password security research community. 
However, we deliberately avoid releasing any sensitive datasets 
or trained models that could potentially facilitate malicious 
password cracking attempts. The released code is accompanied by 
strict usage guidelines, and we explicitly require that any 
applications of our code adhere to responsible research and 
security-oriented purposes only.

%% file: appendix.tex
\begin{table*}[htbp]
  \centering
  \caption{Silhouette scores for different training datasets evaluated with cluster numbers ranging from 30 to 70. 
  Bold values indicate the silhouette scores corresponding 
  to our selected number of clusters for each dataset.}
  \label{tab:silhouette-scores}
  \begin{tabular}{c|ccccccccc}
    \toprule
    Dataset & 30 & 35 & 40 & 45 & 50 & 55 & 60 & 65 & 70 \\
    \midrule
    \texttt{CSDN} &
    0.6046 & 0.6672 & 0.6777 & 0.6837 & \textbf{0.7072} & 0.7083 & 0.7168 & 0.7362 & 0.7783 \\

    \texttt{178} &
    0.6950 & \textbf{0.7072} & 0.7666 & 0.8129 & 0.8158 & 0.8169 & 0.8376 & 0.8611 & 0.8752 \\
    
    \texttt{Rockyou} &
    0.5945 & 0.6195 & 0.6695 & 0.6943 & 0.6978 & \textbf{0.7428} & 0.7725 & 0.7703 & 0.8008 \\

    \texttt{Collection\#1} (processed)&
    0.4061 & 0.4010 & 0.4176 & 0.4298 & 0.4453 & 0.4668 & 0.4812 & \textbf{0.5182} & 0.5137 \\
    
    \bottomrule
  \end{tabular}
\end{table*}

\section{A Proof for Effectiveness of \name}
\label{sec:proof}

\noindent\textbf{Definitions: }
\begin{itemize}
    \item \( p(y) \): The true probability distribution of passwords.
    \item \( \hat{p}(y | \theta) \): The estimated probability distribution of passwords by a single global model with parameters \( \theta \).
    \item \( K \): The number of clusters in the passwords.
    \item \( p_k(y) \): The true subdistribution of passwords in the \( k \)-th cluster.
    \item \( \pi_k \): The prior probability of the \( k \)-th cluster, where \( \sum_{k=1}^{K} \pi_k = 1 \).
    \item \( p_k(y | \theta_k) \): The estimated subdistribution of passwords in the \( k \)-th cluster by the \( k \)-th expert model with parameters \( \theta_k \).
    \item \( w_k \): The weight assigned to the \( k \)-th expert model by the gating mechanism, where \( \sum_{k=1}^{K} w_k = 1 \).
    \item \( \hat{p}(y) \): The estimated probability distribution of passwords by the Mixture of Password Experts (\name) model, defined as \( \hat{p}(y) = \sum_{k=1}^{K} w_k \cdot p_k(y | \theta_k) \).
    \item \( \text{KL}(p \parallel q) \): The Kullback-Leibler (KL) divergence between distributions \( p \) and \( q \), given by \( \mathbb{E}_{y \sim p} \left[ \log \frac{p(y)}{q(y)} \right] \).
    \item \( \text{Error}_{\text{MoPE}} \): The modeling error of \name, defined as \( \text{KL}(p(y) \parallel \hat{p}(y)) \).
    \item \( \text{Error}_{\text{global}} \): The modeling error of the single global model, defined as \( \text{KL}(p(y) \parallel \hat{p}(y | \theta)) \).
\end{itemize}

\noindent\textbf{Objective: }
The goal is to demonstrate that the modeling error of \name, quantified by the Kullback-Leibler (KL) divergence, is significantly lower than that of a traditional single model. Specifically, we aim to prove:
\[
\text{Error}_{\text{MoPE}} \ll \text{Error}_{\text{global}},
\]
where:
\begin{itemize}
    \item $\text{Error}_{\text{MoPE}} = \text{KL}(p(y) \parallel \hat{p}(y))$ is the modeling error of \name,
    \item $\text{Error}_{\text{global}} = \text{KL}(p(y) \parallel \hat{p}(y | \theta))$ is the modeling error of a single global model with parameters $\theta$.
\end{itemize}

\noindent\textbf{Assumptions: }
\begin{enumerate}
    \item \textbf{Clustering Quality is High}: The password data is divided into $K$ clusters, each corresponding to a simpler subdistribution of passwords. The true distribution $p(y)$ can be expressed as a mixture of subdistributions:
    \[
    p(y) = \sum_{k=1}^{K} \pi_k p_k(y),
    \]
    where $p_k(y)$ is the true subdistribution of the $k$-th cluster, and $\pi_k$ is the prior probability of the $k$-th cluster, satisfying $\sum_{k=1}^{K} \pi_k = 1$. 

    Due to the simplicity of each subdistribution, each expert model fits its corresponding subdistribution more accurately than a single model fits the entire distribution, i.e.,
    \[
    \text{KL}(p_k(y) \parallel p_k(y | \theta_k)) \ll \text{KL}(p(y) \parallel \hat{p}(y | \theta)). 
    \]
    If the subdistributions are similar to the overall distribution, then the two KL divergences are approximately equal, i.e.,
  \[
  \text{KL}(p_k(y) \parallel p_k(y | \theta_k)) \approx \text{KL}(p(y) \parallel \hat{p}(y | \theta)).
  \]

    \item \textbf{Gate Method is Accurate}: The gate method in \name assigns weights $w_k$ to each expert model based on input features, 
    such that $w_k \approx \pi_k$ for each cluster $k$.
\end{enumerate}

\noindent\textbf{Proof: }
For a traditional single model (e.g., FLA or PassBERT), the estimated distribution is $\hat{p}(y | \theta)$, and the modeling error is:
\[
\text{Error}_{\text{global}} = \text{KL}(p(y) \parallel \hat{p}(y | \theta)) = \mathbb{E}_{y \sim p(y)} \left[ \log \frac{p(y)}{\hat{p}(y | \theta)} \right].
\]
Since $p(y)$ is a complex mixture of diverse password patterns, a single model struggles to fit the entire distribution accurately, resulting in a large $\text{Error}_{\text{global}}$.

In \name, password data is divided into $K$ clusters, each modeled by a dedicated expert model. The estimated distribution is:
\[
\hat{p}(y) = \sum_{k=1}^{K} w_k \cdot p_k(y | \theta_k),
\]
where $p_k(y | \theta_k)$ is the $k$-th expert model's estimate of the subdistribution $p_k(y)$, $\theta_k$ are the parameters of the $k$-th expert, and $w_k$ are weights from the gating mechanism, satisfying $\sum_{k=1}^{K} w_k = 1$. The modeling error is:
\[
\text{Error}_{\text{MoPE}} = \text{KL}(p(y) \parallel \hat{p}(y)) = \mathbb{E}_{y \sim p(y)} \left[ \log \frac{p(y)}{\hat{p}(y)} \right].
\]

Given the assumption that the gate method is accurate ($w_k \approx \pi_k$), we approximate the estimated distribution as:
\[
\hat{p}(y) \approx \sum_{k=1}^{K} \pi_k \cdot p_k(y | \theta_k).
\]
Thus, the KL divergence becomes:
\[
\text{Error}_{\text{MoPE}} \approx \text{KL}\left( \sum_{k=1}^{K} \pi_k p_k(y) \parallel \sum_{k=1}^{K} \pi_k p_k(y | \theta_k) \right).
\]
Using the convexity of the KL divergence, we obtain the upper bound:

\begin{align*}
  \text{Error}_{\text{MoPE}} & \approx \text{KL}\left( \sum_{k=1}^{K} \pi_k p_k(y) \parallel \sum_{k=1}^{K} \pi_k p_k(y | \theta_k) \right) \\
  &\leq \sum_{k=1}^{K} \pi_k \cdot \text{KL}(p_k(y) \parallel p_k(y | \theta_k)).
\end{align*}

Therefore:
\[
\text{Error}_{\text{MoPE}} \leq \sum_{k=1}^{K} \pi_k \cdot \text{KL}(p_k(y) \parallel p_k(y | \theta_k)).
\]
From the first assumption, each expert's error is significantly smaller than the global error:
\[
\text{KL}(p_k(y) \parallel p_k(y | \theta_k)) \ll \text{KL}(p(y) \parallel \hat{p}(y | \theta)).
\]
Since $\sum_{k=1}^{K} \pi_k = 1$, the weighted sum of these smaller errors is also much smaller than the global error:

\begin{align*}
  \text{Error}_{\text{MoPE}} &\leq \sum_{k=1}^{K} \pi_k \cdot \text{KL}(p_k(y) \parallel p_k(y | \theta_k))   \\
  &\ll \text{KL}(p(y) \parallel \hat{p}(y | \theta)) = \text{Error}_{\text{global}}.
\end{align*}

Thus:
\[
\text{Error}_{\text{MoPE}} \ll \text{Error}_{\text{global}}.
\]

\noindent\textbf{Conclusion: }
Under the assumptions of high clustering quality and accurate gate method, 
we prove that \name achieves a lower modeling error than a traditional single model, especially in scenarios with heterogeneous password data.

\section{Silhouette Score}
\label{sec:sil}

\subsection{Definition}
\label{subsec:sil_define}
The silhouette score~\cite{Rousseeuw1987, wikipedia:silhouette} is a widely used metric for evaluating the quality of clustering results. 
For each data point $i$, the silhouette score $s(i)$ is defined as:
\[
s(i) = \frac{b(i) - a(i)}{\max\{a(i), b(i)\}}
\]
where:

\begin{itemize}
  \item $a(i)$ denotes the average distance between point $i$ and all other points in the same cluster (intra-cluster distance),
  \item $b(i)$ represents the minimum average distance between point $i$ and all points in any other cluster to which $i$ does not belong (nearest-cluster distance).
\end{itemize}

The overall silhouette score for a clustering result with \(k\) clusters is computed as the mean of the individual scores across all data points:
\[
S = \frac{1}{n} \sum_{i=1}^{n} s(i)
\]

where \(s(i)\) is the silhouette score of data point \(i\), and \(n\) is the total number of data points.

The silhouette score \(S\) ranges from -1 to 1 and can be interpreted as follows\cite{wikipedia:silhouette}:

\begin{itemize}
    \item \( S > 0.7 \): Excellent — strong separation and cohesion.
    \item \( 0.5 < S \leq 0.7 \): Good — reasonable separation and cohesion.
    \item \( 0.25 < S \leq 0.5 \): Fair — possible overlap between clusters.
    \item \( S \leq 0.25 \): Poor — potentially meaningless clustering structure.
\end{itemize}

In general, the silhouette score tends to increase with \(k\), 
as more clusters can lead to tighter, more compact groupings. 
However, for \name, a higher silhouette score is not necessarily preferable, 
since it may result in clusters with too few samples, 
which is undesirable for fine-tuning our expert models. 
In extreme cases, if each unique sample is treated as its own cluster, 
the silhouette score would approach 1, but this scenario is clearly impractical.

\subsection{Silhouette Score Result of Training}
\label{subsec:sil_res}
Table~\ref{tab:silhouette-scores} presents the silhouette scores 
used to evaluate the training datasets in Section~\ref{sec:evaluation}, 
which are utilized for training \name. 
We vary the number of clusters from 30 to 70. 
For the three offline training datasets \texttt{CSDN}, \texttt{178}, 
and \texttt{Rockyou}, 
we follow the criterion defined as \textit{Excellent} in Section~\ref{subsec:sil_define}, 
and determine the final number of clusters 
using the method described in Section~\ref{subsec:clustering}. 
For the processed online dataset \texttt{Collection\#1} by edit distance, 
we apply the \textit{Good} criterion to select the number of clusters. 
The silhouette scores in bold in Table~\ref{tab:silhouette-scores} represent 
the scores corresponding to our chosen number of clusters for each dataset. 

It can be observed that when the number of clusters reaches around 60, 
the clustering performance on the offline datasets is evaluated as \textit{excellent}, 
while the performance on the online dataset approaches \textit{good}. 
This indicates that our clustering method introduced in Section~\ref{subsec:clustering} is highly effective.